\documentclass[manuscript,screen,nonacm]{acmart}

\usepackage{amsmath,amsfonts}
\usepackage{array}
\usepackage[caption=false,font=footnotesize,labelfont=sf,textfont=sf]{subfig}
\usepackage{textcomp}
\usepackage{stfloats}
\usepackage{url}
\usepackage{verbatim}
\usepackage{graphicx}
\usepackage{wasysym} 
\usepackage{enumerate}
\usepackage[compact]{titlesec}
\usepackage{enumitem}
\usepackage[ruled,vlined,linesnumbered]{algorithm2e}
\usepackage{graphicx}
\usepackage{caption}
\usepackage{xcolor}
\usepackage{listings, multicol}
\usepackage{filecontents}

\AtBeginDocument{%
  \providecommand\BibTeX{{%
    \normalfont B\kern-0.5em{\scshape i\kern-0.25em b}\kern-0.8em\TeX}}}

\setcopyright{acmlicensed}
\copyrightyear{2018}
\acmYear{2018}
\acmDOI{XXXXXXX.XXXXXXX}

\begin{document}

\title{Secure Ownership Management and Transfer of Consumer Internet of Things Devices with Self-sovereign Identity}

\author{Nazmus Sakib}
\email{nazmus.sakib2411@gmail.com}
\author{Md Yeasin Ali}
\email{yeasin.ali@bracu.ac.bd}
\author{Nuran Mubashshira Momo}
\email{mubashshira.momo@gmail.com}
\author{Marzia Islam Mumu}
\email{marzia.islam.mumu@g.bracu.ac.bd}
\affiliation{%
  \department{Department of Computer Science and Engineering}
  \institution{BRAC University}
  \streetaddress{224 Bir Uttam Rafiqul Islam Avenue}
  \city{Dhaka}
  \country{Bangladesh}}

\author{Masum Al Nahid}
\email{masum@cryptic-consultancy.co.uk}
\author{Fairuz Rahaman Chowdhury}
\email{fairuz@cryptic-consultancy.co.uk}
\affiliation{%
  \institution{Cryptic Consultancy Limited}
  \city{London}
  \country{UK}}

\author{Md Sadek Ferdous}
\affiliation{%
\department{Department of Computer Science and Engineering}
  \institution{BRAC University}
  \streetaddress{224 Bir Uttam Rafiqul Islam Avenue}
  \city{Dhaka}
  \country{Bangladesh}
  }
\affiliation{%
    \institution{ Imperial College London}
     \department{Imperial College Business School}
     \streetaddress{Exhibition Rd} 
     \city{London} 
     \country{UK}
     }
\email{sadek.ferdous@bracu.ac.bd}

\renewcommand{\shortauthors}{N. Sakib et al.}

\begin{abstract}
The popularity of the Internet of Things (IoT) has driven its usage in our homes and industries over the past 10-12 years. However, there have been some major issues related to identity management and ownership transfer involving IoT devices, particularly for consumer IoT devices, e. g. smart appliances such as smart TVs, smart refrigerators, and so on. There have been a few attempts to address this issue; however, user-centric and effective ownership and identity management of IoT devices have not been very successful so far. Recently, blockchain technology has been used to address these issues with limited success. This article presents a Self-sovereign Identity (SSI) based system that facilitates a secure and user-centric ownership management and transfer of consumer IoT devices. The system leverages a number of emerging technologies, such as blockchain and decentralized identifiers (DID), verifiable credentials (VC), under the umbrella of SSI. We present the architecture of the system based on a threat model and requirement analysis, discuss the implementation of a Proof-of-Concept based on the proposed system and illustrate a number of use-cases with their detailed protocol flows. Furthermore, we analyse its security using \textit{ProVerif}, a state-of-the art protocol verification tool and examine its performance.
\end{abstract}



\maketitle

\section{Introduction}
Over time, humans have always sought ways to simplify and control their lives. The Internet of Things (IoT) is a technological advancement that facilitates this desire for ease and control. IoT devices automatically collect data from the environment using sensors, process it, and transmit it to designated recipients for communication or desired outcomes. Originally used for industrial purposes, IoT devices have now been extended to the consumer sector, with the aim of making home life more convenient \cite{rock2022usage}. The growing interest in IoT devices is evident from market demand and capitalisation statistics, highlighting its increasing popularity as a means to lead a more convenient and connected life. 

Based on the Global IoT end-user spending worldwide from 2017 to 2025, the market size has witnessed a significant growth. In 2017, it stood at 110 billion dollars and is projected to reach 800 billion dollars in 2023 and 1,567 billion dollars in 2025 \cite{pv-2}. Furthermore, the Global Consumer IoT Market Size analysis reveals that the market is projected to reach 153.80 billion dollars in 2028, compared to 44.46 billion dollars in 2020 \cite{pv-4}. In general, these statistics demonstrate the increasing interest in consumer IoT devices, with people embracing the convenience and possibilities they offer for a more streamlined and automated way of living and a connected lifestyle. The prominent examples of such consumer IoT devices are smart appliances such as smart refrigerators, smart lights, smart plugs, smart thermostats and so on. In addition, smart kitchen appliances, such as smart ovens and entertainment devices such as smart TVs, smart speakers, and voice-activated assisting devices are also available on the market.

Like legacy home consumer devices, e.g. TVs and refrigerators, users would buy and sell used smart consumer IoT devices. Even though these devices have smart capabilities, their buying and selling take place in a traditional way: the previous owner just hands over the receipt of the product to the new owner, and whoever has the receipt of the respective device can claim to be the owner of the device. We argue that the ownership management of these smart devices in this traditional way could be improved by taking advantage of the smart capabilities of these consumer IoT devices. Here, the term \textit{ownership} refers to ``the ability to control, manage and access a particular device” \cite{sc4}. Unfortunately, there remains a significant gap when ownership management is handled in the traditional way as explained below.

It is imperative that an owner authenticates to the respective IoT device to control, manage and access it. Currently, many smart consumer IoT devices on the market require a smartphone for configuration, management, or control through a smartphone application provided by the OEM (Original Equipment Manufacturer) of the device \cite{16_of_sc5}. Selling of devices, theft, shifting houses or even lending to other owners can accidentally or unintentionally leak data. For instance, when selling smart TV boxes, the new owner may scrap the former owner's data, which even may include credit card information, if the ownership transfer is not handled in a proper way or users fail to log out from the respective device. Therefore, a secure ownership transfer mechanism is necessary considering the modern IoT perspective. Therefore, we argue that, with the proliferation of advanced controllable smart features and the growing number of users, IoT requires identity management that also considers the concept of ownership management into the IoT industry. 

The traditional Silo model \cite{josang2007usability}  requires manufacturers to assign digital identity credentials to users, allowing access to devices. However, this approach presents challenges in terms of managing multiple credentials and securely transferring ownership. In response, the Federated model emerged, outsourcing identity management to third-party entities and leveraging popular login options like ``Login with Facebook'' or ``Login with Google.'' While the Federated model offers convenience, it raises concerns regarding privacy, as user data is consolidated by large corporations.

In recent years, blockchain has gained popularity as a potential component for addressing numerous IoT problems. There are many research to integrate blockchain in IoT, and even an evaluation framework has been proposed to choose a platform for different IoT applications \cite{sdf-iot-survey}. Additionally, the emergence of blockchain technology has also provided an alternative approach to address the shortcomings of various traditional identity models \textcolor{blue}{\cite{shuhan2023decentralised, el2019analysis}}. Blockchain, with its decentralised and immutable nature, offers a promising solution for identity management. Self-Sovereign Identity (SSI) has emerged as a user-centric approach within the blockchain ecosystem \cite{preukschat2021self}. It empowers individuals to create and control their digital identities throughout its life-cycle leveraging decentralized identifiers (DIDs) and verifiable credentials (VCs)\cite{ferdous2019search}. Users can maintain their identities in digital wallets and selectively disclose information as needed, giving them autonomy and privacy in managing their personal data.

By adopting Self-sovereign Identity and leveraging blockchain technology, the IoT industry can move towards a more secure and user-centric approach to ownership transfer and identity management. This paradigm shift has the potential to enhance security, privacy, and trust in IoT ecosystems, empowering users with greater control over their devices and personal data. 

In this article, in order to address the identified issues, we present an SSI and blockchain based holistic solution for secure ownership management and transfer of consumer IoT devices. The major contributions of this article are:
\begin{itemize}
    \item We present a proposal of a holistic SSI-based ownership transfer and management system for consumer IoT devices.
    \item We present its architecture based on a threat model and requirement analysis. 
    \item We discuss different aspects of the Proof-of-Concept (PoC) implementation based on the presented architecture.
    \item We outline different use-cases using the PoC to show the applicability of the proposal.
    \item We analyse its performance and examine the security of the protocol using \textit{ProVerif} \cite{blanchet2018proverif}, a state-of-the-art Protocol Verifier.
\end{itemize}

\textbf{Structure:} We present the relevant background in Section \ref{section:background}. Related research works are reviewed in Section \ref{section:literature-review}. Next, different aspects of the proposed system such as threat model, different requirements ,architecture and implementation are described in Section \ref{section:proposal}. Section \ref{section:proto} outlines different use-cases with detailed protocol flows. We analyse the performance of the solution in Section \ref{section:performance-evaluation}. In Section \ref{section:discussion}, we thoroughly discuss how the proposed mechanism satisfies different requirements, validate the protocol, highlight its advantages and limitation, present a comparative analysis, and describe possible future works. Finally, we conclude in Section \ref{section:conclusion}.

\section{Background}
\label{section:background}
In this section, we provide a brief background on consumer IoT devices (Section \ref{subsection:iot-devices}), blockchain (Section \ref{subsection:blockchain}) and SSI (Section \ref{subsection:ssi}). 

\subsection{Consumer Internet of Things (IoT) Devices}
\label{subsection:iot-devices}
IoT devices are devices designed to automate many tasks. There is no need for any human-machine interaction to run these devices. An IoT is essentially a network of interconnected devices that connect and share data with other devices and are typically equipped with software and sensors, as well as mechanical and digital machinery \cite{pv-1}. Sometimes, IoT devices share the sensor data they collect by connecting to an IoT gateway or another edge device. From there, the data are either sent to the cloud for analysis or analysed locally. These devices can talk to each other and act on the basis of the information they receive from each other. Most of the work is done by the devices without any human intervention; however, owners can set them up, give them instructions, access the data or even change the ownership. The way these Internet-enabled devices connect, network, and talk to each other depends a lot on the IoT applications. 

The Internet of Things has numerous real-world applications, ranging from consumer IoT and enterprise IoT to manufacturing and industrial IoT (IIoT). IoT applications span numerous industries, such as automotive, healthcare, communications and energy \cite{nivzetic2020internet}. There are different mechanisms to control these devices: i) they can be remotely managed and controlled from a smartphone/web application via wi-fi, or cloud server (e.g. smart lock and smart switch) and ii) they can be physically controlled which is directly controlled from the physical interface of the device(e.g. smart card based door lock ).

In the consumer market, smart homes equipped with smart thermostats, smart appliances and connected heating, lighting, and numerous electronic devices can be remotely controlled via computers and smartphones. Wearable devices equipped with sensors and software can collect and analyze user data, sending messages to other technologies about users in an effort to make their lives more convenient and comfortable. Wearable devices are also used for public safety, for instance, to improve the response times of first responders during emergencies by providing optimized routes to a location or by monitoring vital signs of construction workers or firefighters at life-threatening sites. IoT offers numerous advantages in healthcare, including the ability to monitor patients more closely by analyzing generated data. Hospitals often employ IoT systems for tasks such as inventory management of pharmaceuticals and medical instruments. Using sensors to detect the number of occupants in a room, intelligent buildings can reduce energy costs, for example. The temperature can be adjusted automatically, for instance, by turning on the air conditioner when sensors detect a full conference room or turning down the heat when everyone has left for the day. Using connected sensors, IoT-based smart farming systems can monitor, for example, the light, temperature, humidity, and soil moisture of crop fields. IoT also contributes to the automation of irrigation systems. IoT sensors and deployments, such as smart streetlights and smart meters, can reduce traffic, conserve energy, monitor and address environmental concerns and enhance sanitation in a smart city \cite{pv-1}.

\subsection{Blockchain}
\label{subsection:blockchain}
With the introduction of Bitcoin \cite{nakamoto2008bitcoin}, a new technology called blockchain emerged. A blockchain is a decentralized, immutable, transparent, and append-only ledger system that shares and validates resources through distributed peer-to-peer nodes \cite{j-morshed}. It is a linked data structure forming a sequential chain of blocks, each containing a collection of transactions. These transactions log actions for processing or transferring currency or data. Blockchain relies on cryptography for validation and the creation of new blocks. Each new block is appended to the previous one, referring to it via cryptographic mechanisms, thus creating the blockchain. This data structure becomes immutable once formed. Consensus algorithms such as Proof of Work (PoW), Proof of Stake (PoS), Proof of Authority (PoA), Proof of Elapsed Time (PoET), and Practical Byzantine Fault Tolerant (PBFT) are used to ensure network-wide agreement on transaction and block orders \cite{ferdous2020blockchain}.

Following Bitcoin, Ethereum introduced the notion of 'smart-contracts' to blockchain technology \cite{smart-contract}. This development transformed blockchain into a distributed computing system with a trusted source of truth and autonomous code execution. As a result, various applications have been developed in finance, supply chain, IoT, land registration, healthcare and so on \cite{public-blockchain-usecase}. 

A blockchain system can be public, meaning it is open to all and anyone can join the network and carry out transactions. Additionally, a blockchain system can be private (also known as permissioned) where only the authorised entities can participate. Bitcoin \cite{bitcoin-org}, Ethereum \cite{Ethereum} and Solana \cite{solana} are examples of some popular public blockchains. On the other hand, Hyperledger projects \cite{hyperledger-projects}, Ethereum enterprise \cite{ethereum-enterprise}, R3 Corda \cite{r3-corda} are examples of some permissioned blockchain systems. 

\subsection{Self-sovereign Identity (SSI)}
\label{subsection:ssi}
Identity management is an indispensable component of any online service. The online identity is the information used to verify the users of the system. It helps authenticate users to the system and allows them to access the system to access its services. A system that is used for identity management is known as an Identity Management System (IMS). With the evolution of online services, there are many different types of identity models and systems have been introduced. There are a number of identity management models, however, a few popular ones are the SILO model and the Federated Identity model \cite{josang2005user, ferdous2015user}. 

Even though these two models are quite popular, they suffer from serious limitations. They are centralised in nature exhibiting a single point of failure and offer limited control over the identity of the users \cite{ferdous2015user}. Once identity data are stored, users have minimal control over preventing third-party access to their data. Ultimately, users are unaware of how providers might misuse their data.

In order to address the issues of existing identity models, a new identity model called \textit{Self-sovereign Identity (SSI)} has been introduced. Its main motivation is to empower individuals by granting them greater control over their identity data. It represents a new paradigm in managing digital identities, where the user creates and maintains control throughout its lifecycle \cite{ferdous2019search, muehle_ssi}. To realise the notion of SSI, the World Wide Web Consortium (W3C) developed a number of key technologies such as \textit{Decentralized Identities} (DIDs) \cite{w3cdids} and \textit{Verifiable Credentials} (VCs) \cite{w3cvcs} as central constructs for SSI. 

There are three different entities within SSI: users (subjects or holders), issuers and verifiers. DIDs are unique identifiers which are associated with the cryptographic public key of an SSI entity. DID Documents, or DID Docs, are JSON objects that include the DID, its associated cryptographic public keys, and additional metadata. Conversely, (VCs) are cryptographically-signed claims (consisting of attribute names and their respective values) about a subject made by an issuer. Because VCs are signed by issuers, their authenticity can be confirmed by any verifier who retrieves the corresponding DID and validates the signature using the associated public key. The owner of these VCs keeps them in a digital wallet and can present them, either singly or in combination, to a verifier in exchange for a favourable action, such as access to a service. In this context, blockchain serves as a verifiable data registry (VDR) for storing DIDs or DID Docs because of its data immutability and decentralisation characteristics. There is another optional entity, called a \textit{Mediator}, within the SSI ecosystem. The responsibility of a mediator is to forward messages between a holder and an issuer/verifier. In addition, the mediator holds messages for an offline entity so that it can deliver messages later when the entity becomes online.

In order to engage in the aforementioned interactions, the SSI entities need to participate in the following steps (Figure \ref{Fig:SSI}):

\begin{figure}[ht]
    \centering
    \includegraphics[width=0.65 \linewidth]{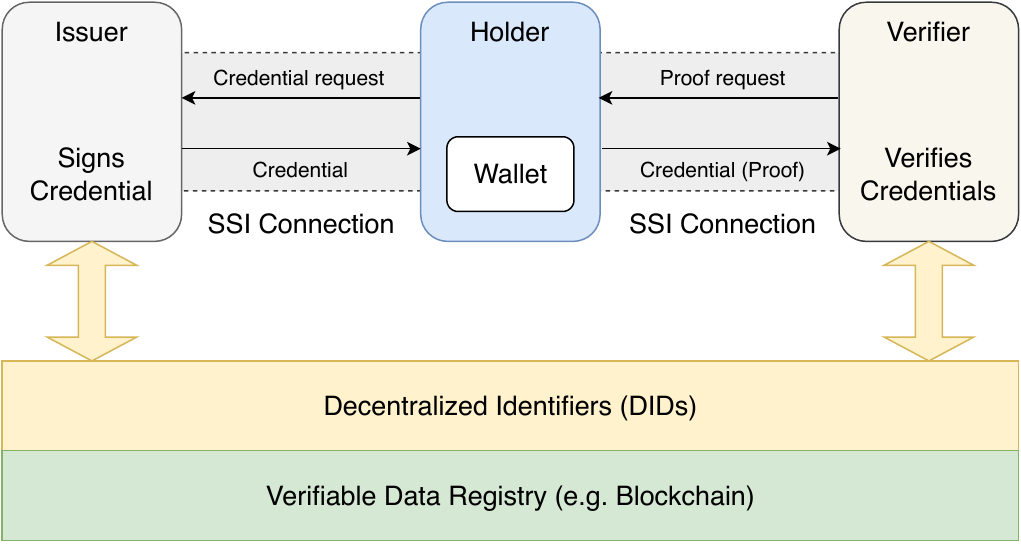}
    \caption{SSI entities and their relations \cite{preukschat2021self}.} 
    \label{Fig:SSI}     
\end{figure}   

\begin{itemize}
    \item \textbf{Establishing an SSI connection:} 
    At the beginning, any two SSI entities (an issuer and a holder, or a holder and a verifier) need to establish a pairwise SSI connection. This connection is encrypted to guarantee the confidentiality of the data exchanged using the established SSI connection.    
    \item \textbf{Providing a VC:} Upon the user's request, the issuer generates a VC and transmits it through the already established SSI connection.    
    \item \textbf{VC storage:} The user wallet receives this VC and validates it by verifying the digital signature on the VC. Then, the VC is stored in the wallet.
    \item \textbf{Requesting proof:} 
    When the user attempts to access a service from the verifier (assuming the verifier serves as a Service Provider, SP), the verifier requests proof from the user. The user then submits the VC via the pre-established SSI connection with the verifier. The verifier then checks the digital signature on the presented VC and makes an authorisation decision to permit or deny access to the user.
\end{itemize}

\section{Literature Review}
\label{section:literature-review}
In this section, we present a summary of relevant research works. The concept of ownership transfer using RFID (Radio Frequency Identification) tag first presented by Molnar et el. with a proposed protocol in 2005 \cite{first_ownership1}. With the advancement of IoT, research on ownership transfer of RFID-based IoT devices has seen a lot of exposure. From centralised models \cite{RFID_SC1_CENTRALIZE,RFID_SC2_CENTRALIZE} to decentralised schemes \cite{RFID_SC3_DECENTRALIZE_1, RFID_DECENTRALIZE_2}, RFID-based ownership transfer in IoT covered a lot of areas in IoT particularly when a large number of products needed to be tracked, e.g. in supply chain, inventory,  production management and manufacturing \cite{ HC3, RFID_INVENTORY_1, RFID_NON_IOT_OWNERSHIP_TRANSFER, CHALLENGE_DISTANCE}. However, One of the limitations of RFID-based IoT is that the RFID tag requires an RFID reader and the device should be placed within a short distance so that the reader can read the tag \cite{CHALLENGE_DISTANCE}. Therefore, this type of feature is not suitable for devices that are remotely controlled by smartphone or web applications. 

Vidyadhar Aski et al.\cite{sc1} proposed a framework that focused on the access control and ownership transfer of medical IoT devices for edge computing-based applications. The proposed architecture was divided into three entity categories such as patient, healthcare professional and healthcare server and there were five phases to use it. Although the authors presented multiple functionalities, the ownership transfer phase was their major contribution. According to the transfer mechanism, users are required to use a smart card, credentials (user ID, password) and biometric fingerprint. The paper presented some security analysis and protocol requirements, but there was no protocol design, implementation, and system validation.

S.F. Aghili et al. \cite{sc5} proposed a scheme for an IoT-based e-health system. One of its contributions was ownership transfer from the perspective of an e-health system using a medical server. According to the proposed scheme, healthcare professionals could transfer ownership access to patient data from one sensor to another. For the process, the scheme was dependent on a physical smartcard. In addition, biometric information was also required along with user ID and password. This work had a detailed protocol design and validation, as well as security and functional analysis. However, its communication channel was insecure.

Gunnarsson et al. \cite{sc2} proposed an ownership transfer protocol that utilised a symmetric key on the IoT side considering the resource constraints of IoT environments. The protocol included the deployment, ownership transfer preparation and finally the ownership transfer phases. However, their proposal is partially dependent on a trusted third party (TTP) node which they referred to as the ``Reset Server (RS)''. However, unlike other TTPs, RS does not know the device credentials after the transfer process. Although the paper has many technical and implementation details, it focuses on the whole infrastructure of products instead of a single product. 

To achieve an automatic authentication and transfer process, Khan et al. \cite{sc4} proposed a system called ChownIoT that combined the authentication of the smartphone and its respective user, profile management, data protection, and an automated ownership change process. In the research, the ownership change context referred only to the IoT device. The main goal of ChownIoT is to protect user-credential data on the device and any associated data stored in the cloud. ChownIoT used contexts such as the network where it was deployed, and upon a change of context, it automatically initiated the authentication process and if required, initiated the ownership change process. The paper focused on ownership change from the device’s perspective and implemented a PoC (Proof-of-Concept), there was no detailed protocol design.

X. Leng et al. \cite{pv-9} proposed an idea of ownership transfer based on a TTP. In this proposal, owner registration and the establishment of the security credentials with the particular IoT device were required. Moreover, the TTP was responsible for generating ownership challenges. Furthermore, the IoT device had a pre-installed secret shared with a TTP and had a one-way function to generate the secret key for ownership. An ownership creation model was used when the IoT device ownership was first registered. The transfer procedure required that both the current owner and the new owner to be registered with the TTP. To transfer ownership, a trust should be established between both owners: the current and new owner which included permissions and access control. A framework based on item-level access control through a mutual trust was presented for IoT devices. The main limitation of their proposal was the involvement of a TTP, creating a notion of a single-point of failure.

B.A. Barakati et al. \cite{HC1} proposed an approach which they defined as IoT of Trust that focused to eliminate a central authority by adopting blockchain for its decentralised architecture based on Ethereum. They also defined a definition for ownership management in the context of IoT. Although they implemented a PoC, the theoretical and PoC presentations lacked technical details.

M. Alblooshi et al. \cite{HC2} proposed a similar system using Ethereum for the management of device ownership. They divided the smart-contract in Ethereum into two contracts: one is used by the Manufacturer for the device when it is first created, and the other one is used by the owner for maintaining the ownership. However, it does not present any protocol flow of the system or any technical evaluation of the system.

Leveraging Physically Unclonable Function (PUF) and blockchain together, Kiran M. et.al \cite{New-1} proposed a protocol to transfer ownership which motivated from the issues in a supply chain management. In their scenario, both the device and the parties involved in the supply chain management needed to be authenticated. Their proposal also introduced partial ownership transfer and presented the verification of their protocol, however, they did not use any well-established threat model framework. Moreover, there were no specified and structured functional and security requirements.

\section{Proposed System}
\label{section:proposal}
In this section, we present our proposal for an SSI-based ownership transfer management system for consumer IoT devices. Toward this aim, we highlight the important functionalities of the system in Section \ref{subsection:system-proposal}. Next, we discuss the high-level architecture of the system in Section \ref{subsection:architecture-design} and the implementation details in Section \ref{subsection:implementation}.

\subsection{System Proposal}
\label{subsection:system-proposal}
Managing identities is a crucial functionality of any consumer IoT device. This is because any consumer IoT device should have the ability to identify and authenticate their true owner to facilitate ultimate control over the respective device. This will enable the owner to control the device as required. Moreover, when a consumer IoT device is sold to someone else, the respective IoT device should have the knowledge about this transfer of ownership. This will help the device to ensure that any access rights of the previous owner are fully revoked, and the new owner establishes their access control rights to the device. To facilitate this, the IoT device must have the provision of identifying and authenticating the new owner. In this section, we present our proposal on how different use-cases within this setting can be facilitated using SSI. The proposal is holistic in nature in the sense that it engages all possible entities within the ecosystem such as the manufacturer, seller and multiple buyers. These entities assume different SSI roles, e.g. issuers and verifiers, according to the scenario. Since the proposed system will deal with sensitive user information, it is important to consider different security and privacy threats. Therefore, we present a threat model for the proposed system in Section \ref{subsection:threat-modelling}. Then, we discuss different functional, security and privacy requirements for the proposed system in Section \ref{section:requirements-analysis}. 

\subsubsection{Threat modelling}
\label{subsection:threat-modelling}
Threat modelling is an important step toward identifying threats in a secure system. It helps to detect the threat early in the software development life-cycle and helps to formulate the security requirements of an application. For this research, we have chosen STRIDE \cite{shostack2014threat} which is one of the well-established threat modelling frameworks. STRIDE represents an acronym of six threats: Spoofing identity, Tampering, Repudiation, Information disclosure, Denial of Service and Elevation of privilege. Additionally, we have taken into account one additional security: replay threat. In the following, we discuss how these threats apply to our proposed system.

\vspace{2mm}
\noindent \textbf{Security Threats:}

\begin{enumerate}[label=T\arabic*.]
    \item 
\textbf{Spoofing Identity:} Spoofing identity is pretending to be someone else. A spoofing attack arising from this threat allows an individual (referred to as a \textit{Spoofer}) to impersonate another person, deceiving the target into divulging their personal information or performing actions for the attacker's benefit. Within the scope of this research, this threats implies that an Spoofer can access the manufacturer's services spoofing someone else's identity for claiming or proving the ownership of an IoT device.

\item \textbf{Tampering Data:} Data tampering occurs when data or information is altered without proper authorisation. This risk suggests that an attacker might manipulate VC, PIN, or other essential data with malicious intent.

\item \textbf{Repudiation:} The primary concern of a repudiation threat involves an actor denying any wrongful or damaging activities they have conducted, such as denying different actions while claiming or transferring ownership.

\item \textbf{Information Disclosure:} Information disclosure threats occur when sensitive information is inadvertently exposed to attackers or unauthorised individuals. 
 
\item \textbf{Denial-of-Service (DoS):} The goal of a Denial-of-Service (DoS) attack is to slow or shut down a network or servers so that its users cannot use it. With respect to the scope of our research work, DoS will be relevant to the online services which will be used for the management and transfer of ownership information. 
 
\item \textbf{Elevation of Privilege Threats:} Elevation of privilege is the threat where an attacker uses different attack vectors to elevate its access privilege of different services of the proposed system.

\item \textbf{Reply Attack:} This attack implies that an attacker can capture a request or response and reuse it later for a malicious purpose.
\end{enumerate}

\vspace{2mm}
\noindent \textbf{Privacy Threats:}
In addition to different security threats, we also consider the following privacy threats.
\begin{enumerate}[label=T\arabic*., start = 8]  
    \item \textbf{Lack of consent:} This implies that important data are shared with someone without the consent of the respective holder.
\end{enumerate}


\subsubsection{Requirement Analysis}
\label{section:requirements-analysis}

In this section, we present different functional , security and privacy requirements. Functional requirements (FR) represent the core functionalities of the system that the system must have. On the other hand, the security requirements (SR) and privacy requirements (PR) represent the requirements which can be used to mitigate the identified security and privacy threats respectively.  we present the security(SR) and privacy(PR) requirements.

\vspace{2mm}
\noindent \textbf{Functional Requirements:} The functional requirements of the system are presented below:

\begin{enumerate}[label=FR\arabic*.]    
    \item The system must adopt the SSI mechanism to facilitate the ownership transfer and management of IoT devices. 
    \item In order to accommodate SSI functionalities, the system must adjust the roles of existing entities to make them functional as SSI entities. 
    \item The system should change how ownership information is represented, attached to a specific IoT device and maintained so that it is SSI compliant.
    \item Once an IoT device is sold, the system must ensure that the ownership of the previous owner is properly revoked and the ownership of the new owner is instilled. 
\end{enumerate}
\vspace{2mm}
\noindent \textbf{Security Requirements:} The security requirements for the proposed system are discussed next. 
\begin{enumerate}[label=SR\arabic*.]
    \item The proposed system must ensure that only the real owner can request the manufacturers to transfer their ownership and that the real buyer can claim that ownership. Therefore, the system must verify the identity of the owners and buyers in a way that is hard to spoof. This will mitigate the $T1$ threat. 
    \item The system must have the mechanism to prevent the modification of important data, particularly the ownership data and PIN. This will mitigate the $T2$ threat. 
    \item Crucial data must be digitally signed by the respective entity to mitigate threat $T3$.
    \item The system should use encrypted channels to ensure that unauthorised entities cannot access the data while they are in transit. Similarly, important data must be kept in encrypted format when stored in a database. This will mitigate the $T4$ threat.
    \item The system should ensure precautionary steps to protect the relevant services against any DoS attack, thus mitigating the threat $T5$.
    \item The system should adopt the necessary access control mechanisms to prevent any access privilege elevation to mitigate $T6$.
    \item The system must ensure precautionary steps to protect against any reply attack, thus mitigating $T7$.
\end{enumerate}

\vspace{2mm}
\noindent \textbf{Privacy Requirements:}
The privacy requirements of the proposed system are presented below:
\begin{enumerate}[label=PR\arabic*.]
    \item Every important datum should be shared only after the respective owner has consented to do so. This mitigates $T8$.
\end{enumerate}

\subsection{Architecture}
\label{subsection:architecture-design}
In the proposed system is holistic in nature in the sense that it considers the majority of the entities which are crucial in the whole ecosystem of the application domain. The considered entities of the system are: i) Manufacturers, ii) Distributors, iii) Buyers and iv) Sellers. The manufacturer ($MF$) is the original equipment manufacturer (OEM) that produces consumer IoT devices. According to our design, manufacturers do not sell products directly to end users. The \textit{Distributor ($DS$)} is the entity that sells products to end users. In the real world, there are selling agents, retailers, dealers and other intermediaries that sell products. However, for simplicity, we denote all of them as distributors. We assume that a distributor collects products directly from the manufacturer and sells them to end-users. An end user can be a buyer when the user buys the device from the distributor or from another seller. A seller is an end-user who sells the devices to another end-user. These entities assume different SSI roles depending on the scenarios. For example, a manufacturer assumes the role of an issuer, a distributor assumes the role of a verifier, a buyer assumes the role of an issuer whereas a seller takes the role of a verifier.

\begin{figure}[htbp]
\centering
\includegraphics[scale=0.7]{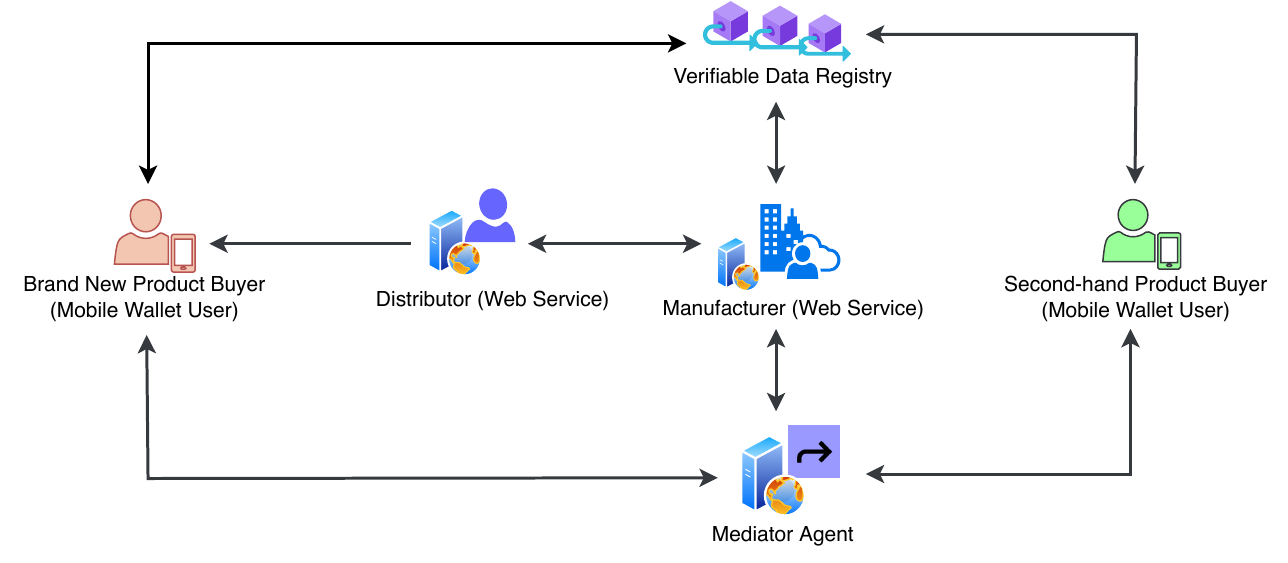}
\caption{Architecture of the proposed system}
\label{fig:architecture}
\end{figure}

Based on these assumptions, the architecture of the proposed system is illustrated in Figure \ref{fig:architecture}. There are a few entities in this architecture. \textit{Manufacturer(MF)} is the entity with which an SSI software agent is integrated to achieve the SSI functionalities. The manufacturer runs a web service that is used to verify the authenticity of an IoT device. An SSI agent is integrated with this web service to facilitate SSI based communications. The distributor also offers a web service whose functionality will be explained later. A buyer would need to establish an SSI communication using an SSI wallet with the manufacturer or another end-user seller to ensure SSI interactions. A buyer/seller uses the services of a mediator to communicate with the manufacturer or another SSI entity. 

Using this architecture, we propose that the ownership information and other required data are represented as VC. A manufacturer issues a VC when a buyer buys an IoT device for the first time using a distributor. This VC is stored in the buyer's wallet. Whenever the previous owner (acting as a seller) sells the IoT device to another user (acting as a buyer), the seller issues another VC to the new owner. All these entities engage in a secure protocol that is discussed in Section \ref{section:proto}.

\subsection{Implementation}
\label{subsection:implementation}
Following the designed architecture of the system, we have developed a Proof-of-Concept (PoC) using the following frameworks and libraries.

\begin{itemize}
    \item The Hyperledger Aries \cite{HyperledgerAries} framework provides a python-based SDK called Hyperledger Aries Cloud Agent Python (ACA-Py, in short) \cite{acapy}. This has been used to facilitate SSI functionalities. Aca-Py internally relies on Hyperledger Indy \cite{HyperledgerIndy} as the VDR. We have adopted Indy testbed called \textit{BCovrin dev} \footnote{http://dev.greenlight.bcovrin.vonx.io} for the PoC.
    \item Aries Mobile Agent React Native (Aries Bifold) \cite{bifold} is the SSI mobile wallet used. We have modified this wallet to add additional features on top of it.
    \item To build the web services of the Manufacturer and the Distributor, NodeJs \cite{nodejs} has been used. SSI functionalities have been integrated with the web services using the ACA-Py SDK. 
    \item An ACA-Py based public mediator called \textit{Indicio Public Mediator} \cite{indicio} has been  used for the PoC.
\end{itemize}

\section{Protocol Flow \& Use-cases}
\label{section:proto}
In this section, we explore how the implemented PoC could be utilised in different use-cases for the ownership transfer and management of consumer IoT devices. Towards this aim, Section \ref{subsection:data-model-and-algoritms} includes the data model and algorithms utilised by the PoC and Section \ref{subsection:usecase-and-protocol-flow} sketches out the use-cases along with the underlying protocol flows between different entities. 

\subsection{Data Model and Algorithms}
\label{subsection:data-model-and-algoritms}
In this section, we present the mathematical notations, data model and algorithms used in the protocol flow.

\vspace{1mm}  
\noindent \textbf{Notations:} Different mathematical and cryptographic notations have been employed in the proposed protocol and its data model. For a better understanding, these notations are listed in Table \ref{table:mathematical-notation}. 

\begin{table*}[htbp]
\caption{Mathematical Notations}
\label{table:mathematical-notation}
\centering
\begin{tabular}{m{3cm} | m{7cm}}
\hline
Notations & Descriptions\\
\hline
$DS$ & A distributor \\ 
\hline
$MF$ & A manufacturer\\
\hline
$B_i$ & A buyer where $i \in \mathbb{Z}^+$ \\
\hline
$S_i$ & A seller where $i \in \mathbb{Z}^+$ \\
\hline
MD & Mediator \\
\hline
ConnID & Connection ID \\
\hline
$ VC^{B_i}_{{MF}} $ & A VC issued by $MF$ to ${B_i}$ \\
\hline
$ PIN^{B_1}_{{MF}} $ & A pin of 6-8 alphanumeric characters created by $MF$ for the buyer $B_1$, who buys the brand new product directly from $DS$  \\
\hline
$ PIN^{{MF} }_{B_2} $ & A pin of 6-8 alphanumeric characters created by buyer $B_2$ for $MF$, while buying the product from the seller ($S_1$); where $S_1 = B_1$.  \\
\hline
$ TID^{B_1}_{MF} $ & A tracking ID created by $MF$ for the buyer $B_1$, who buys a brand new product directly from $DS$;  \\
\hline
$ TID^{B_2}_{S_1} $ & A tracking ID created by the seller $S_1$ for the buyer $B_2$; where $ S_1 = B_1$ \\
\hline
$K^X_{Y}$      & Public key of $Y$ for $X$; where $X = B_i \lor X = S_i \lor X = MF \lor X = MD$ and $Y = B_i \lor Y = S_i \lor Y = MF \lor Y = MD$   \\ \hline
$K^{-1|X}_{Y}$      & Private key of $Y$ to be used for $X$; where $X = B_i \lor X = S_i \lor X = MF \lor X = MD$ and $ Y = B_i \lor Y = S_i \lor Y = MF \lor Y = MD$  \\ \hline
$K_{{B_i}MF}$      & A symmetric key generated by $B_i$ for $MF$   \\ \hline
$DID^X_{Y}$      & Decentralised identifier of $Y$ for $X$\\ \hline
$N_i$    & A fresh nonce   \\ \hline
${\{\}}_{K^X_Y}$     & Encryption operation using a public key $K^X_Y$  \\ \hline
${\{\}}_{K^{-1|X}_{Y}}$     & Signature using a private key $K^{-1|X}_{Y}$  \\ \hline
${\{\}}_{K_{XY}}$     & Encryption operation using a symmetric key $K$  \\ \hline
${[...]}_{\mathit{K^X_Y}||\mathit{HTTPS}||\mathit{SMTP} }$     & Communication over an encrypted channel  with either with $K^X_Y$ or $\mathit{HTTPS}$  or \textit{SMTP} protocol  \\ \hline
\end{tabular}
\end{table*}
%

\textbf{Data Model:} For the PoC, we have developed a request and response protocol, where an entity sends a request and the other entity replies with a respective response. There are a number of requests and responses in our system. We denote the set of requests as $req$ which contains seven requests, namely $productSellingReq$, $ownershipClaimReq$, $PINReq$, $ownershipTransferReq$, $ownershipProofReq$, $pinChallengeReq$, and $revokeVC$ that represent the product selling request to a manufacturer $MF$, ownership claim request, PIN request, ownership transfer request, ownership proof request, PIN challenge request and VC revocation notification respectively. There are seven corresponding responses within the response set (denoted as $resp$): $productSellingResp$, $ownershipClaimResp$, $PINResp$, $ownershipTransferResp$, $ownershipProofResp$, $pinChallengeResp$ and $revokeVCResp$. In addition to the request-response model, we have also used a request-response-acknowledgement structure and $ownershipClaimAck$ is the acknowledgement of $ownershipClaimResp$. All of these request, response or acknowledgement actions contain a number of data. The definition of each of these actions along with the additional data model within the respective actions, is presented in Table \ref{table:data-model}.

\begin{table*}[!h]
\caption{Data Model}
\label{table:data-model}
\centering
\begin{tabular}{p{14.5cm}}
\hline
$\mathit{req} \triangleq \langle \mathit{ productSellingReq, ownershipClaimReq, PINReq, ownershipTransferReq, ownershipProofReq, pinChallengeReq, }$ \\
\hspace{10mm} $revokeVC  \rangle$\\ \hline
$\mathit{resp} \triangleq \langle \mathit{productSellingResp, ownershipClaimResp, PINResp, ownershipTransferResp, ownershipProofResp,  }$ \\
 \hspace{10mm} $pinChallengeResp, revokeVCResp \rangle$\\ \hline 

$\mathit{VC}^{B_i}_{{MF}} \triangleq \langle \mathit{  ((a_1, {av}_1), (a_2,{av}_2), ..., (a_n,{av}_n)) } \rangle$ \\ \hline
$\mathit{productAttr} \triangleq \langle \mathit{productCode, distributorID, ConnID, status, previouslySoldCount, firstPurchaseDate, lastPurchaseDate, }$ \\
\hspace{15mm} $email \rangle$\\ \hline
$\mathit{productSellingReq} \triangleq \langle \mathit{ productAttr } \rangle$\\ \hline
$\mathit{productSellingResp} \triangleq \langle \mathit{ TID^{B_1}_{MF} } \rangle$\\ \hline
$\mathit{ownershipClaimReq} \triangleq \langle \mathit{ (( TID^{B_1}_{MF}, PIN^{B_1}_{MF} ) || ( TID^{B_2}_{S_1}, K_{{B_2}MF} )) } \rangle$\\ \hline
$\mathit{ownershipClaimResp} \triangleq \langle \mathit{ VC^{B_i}_{MF} } \rangle$\\ \hline
$\mathit{ownershipClaimAck} \triangleq \langle \mbox{Responding with an acknowledgement status of } ownershipClaimResp   \rangle$\\ \hline
$\mathit{PINReq} \triangleq \langle \mathit{TID^{B_2}_{S_1} } \rangle$ \\ \hline
$\mathit{PINResp} \triangleq \langle \mathit{ \{{PIN}^{MF}_{B_2} \}_{K_{B_2}{MF}}, TID^{B_2}_{S_1} } \rangle$ \\ \hline
$\mathit{ownershipTransferReq} \triangleq \langle \mathit{productCode, \{ PIN^{MF}_{B_2} \}_{K_{{B_2}{MF}}}, TID^{B_2}_{S_1} } \rangle$\\ \hline
$\mathit{ownershipTransferResp} \triangleq \langle \text{ Responding with an acknowledgement status of $ownershipTransferReq$ } \rangle$\\ \hline
$\mathit{ownershipProofReq} \triangleq \langle \mathit{a_1, a_2, ..., a_n} \rangle$ \\ \hline 
$\mathit{ownershipProofResp} \triangleq \langle \mathit{VC}^{B_i}_{MF}, \mathit{metadata} \rangle$ \\ \hline
$\mathit{pinChallengeReq} \triangleq \langle \mathit{ TID^{B_2}_{S_1}, challengeBy, challengeType } \rangle$\\ \hline
$\mathit{pinChallengeResp} \triangleq \langle \mathit{ TID^{B_2}_{S_1}, challengeResult } \rangle$\\ \hline
$\mathit{revokeVC^{B_i}_{MF}} \triangleq \langle \text{ Notify the revocation of $VC^{B_i}_{MF}$ of associated product from $ownershipClaimReq$; where $S_i=B_i$ } \rangle$\\ \hline
$\mathit{revokeVCResp} \triangleq \langle \text{ Acknowledge $revokeVC$ notification } \rangle$\\ \hline
$\mathit{claimantAttribute} \triangleq \langle \mathit{ (productCode, value), (tid, TID^{B_1}_{MF} || TID^{B_2}_{S_1}), (pin, (PIN^{B_1}_{MF} || \{ PIN^{MF}_{B_2} \}_{{B_2}MF} )), (key, K_{{B_2}MF}), } $\\
\hspace{28mm} $(challengeBy, value), (challengeType, value) \rangle$ \\ \hline
$\mathit{claimantAttributeList} \triangleq \langle \mathit{ claimantAttribute_1, claimantAttribute_2, ......., claimantAttribute_n } \rangle$ \\ \hline
$\mathit{ownershipClaimingData} \triangleq \langle \mathit{ (productCode, value), (tid, TID^{B_1}_{MF} || TID^{B_2}_{S_1}), (pin, PIN^{B_1}_{MF} || PIN^{MF}_{B_2} ), (key, K_{{B_2}MF}), }$ \\ 
\hspace{35mm} $(challengeBy, value), (challengeType, value) \rangle$ \\ \hline
$\mathit{ownershipClaimingDataList} \triangleq \langle \mathit{ ownershipClaimingData_1, ownershipClaimingData_2, ......., ownershipClaimingData_n } \rangle$ \\ \hline
$\mathit{productData} \triangleq \langle \mathit{ productAttr } \rangle$ \\ \hline
$\mathit{productDataList} \triangleq \langle \mathit{ productData_1, productData_2, ......., productData_n } \rangle$ \\ \hline
\end{tabular}
\vspace{-3mm}
\end{table*}


\textbf{Algorithms:} Next, we present some of the important functions of the ownership transfer process in two different algorithm snippets (Algorithm \ref{algo:handle-new-product-ownership-claim} and Algorithm \ref{algo:handle-used-product-ownership-claim}). 


\begin{algorithm}
\SetAlgoLined
\caption{Algorithm Snippet for \texttt{handleNewProductOwnershipClaim}, \texttt{generateVC} \& \texttt{handleOwnershipTransferRequest} functions}
\label{algo:handle-new-product-ownership-claim}
\SetKwBlock{Begin}{}{}
\Begin(\textbf{Start})
{
  $...$\\
  \noindent \tcc{ The function below is handled by $MF$ when $B_1$ requests to claim ownership of a new product.}
    \SetKwFunction{newProductOwnership}{\textbf{handleNewProductOwnershipClaim}}
      \SetKwProg{Fn}{function}{}{}
      \Fn{\newProductOwnership{$\mathit{ownershipClaimReq}$}}{
        $receivedTID$ := $ownershipClaimReq.TID^{B_1}_{MF}$
        $receivedPIN$ := $ownershipClaimReq.PIN^{B_1}_{MF}$ 
        $\mathit{claimantAttribute} := claimantAttributeList.matchAndReturn(receivedTID, receivedPIN)$\;
            \uIf{claimantAttribute == ( Invalid $||$ Empty )}{
                  \KwRet reject $ownershipClaimReq$\;
            }
            \Else{ 
               $productCode := claimantAttribute.productCode$ \\
                $productData := productList.matchAndReturn(productCode)$ \\
                \uIf{$productData$ not found}{
                    \KwRet reject $ownershipClaimReq$\;
                }
                \Else{ 
                 $VC^{B_1}_{MF}$ := $generateVC(productData)$ \\
                 \KwRet $VC^{B_1}_{MF}$
                }
            }
      }
      
  \tcc{ The function below is handled by $MF$ to generate VC.}
  \SetKwFunction{handleVC}{\textbf{generateVC}}
  \SetKwProg{Fn}{function}{}{}
  \Fn{\handleVC{$\mathit{productData}$}}{
    $\mathit{credDefID} := \mbox{Retrieve credential definition ID }$\; 
    $\mathit{VC^{B_1}_{MF}}$ := \mbox{Prepare verifiable credential using $credDefID$ and $productData$ }\; 
    \KwRet $VC^{B_1}_{MF}$ \;
  }

\tcc{ The function below is handled by $MF$ to handle the $ownershipTransferReq$ sent by $B_1$.}
   \SetKwFunction{handleOwnershipTransfer}{\textbf{handleOwnershipTransferRequest}}
      \SetKwProg{Fn}{function}{}{}
      \Fn{\handleOwnershipTransfer{$\mathit{productCode, TID^{B_2}_{S_1}, \{ PIN^{MF}_{B_2} \}_{K_{B_2}{MF}} }$}}{
        $\mathit{isDuplicate}$ := $claimantAttributeList.isDuplicate($productCode$)$ \;
            \uIf{ isDuplicate is TRUE }{
               \KwRet reject $ownershipTransferReq$\;
            }
            \Else{
                Store $\mathit{ productCode, TID^{B_2}_{S_1}, \{ PIN^{MF}_{B_2} \}_{K_{B_2}{MF}} }$ in $claimantAttributeList$ \\
                \KwRet $ownershipProofReq$\;
            }
      }
  ...
}
\end{algorithm}

In Algorithm \ref{algo:handle-new-product-ownership-claim}, all of the functions are handled from the $MF$'s end. The \texttt{handleNewProductOwnershipClaim} function handles the $ownershipClaimReq$ sent from the buyer ($B_1$) to $MF$ to claim the ownership of a brand-new product that the buyer has purchased from a $DS$. The function first extracts $TID^{B_1}_{MF}$ and $PIN^{B_1}_{MF}$ from the buyer $B_1$'s request ($ownershipClaimReq$). Next, the $MF$ verifies the  $TID^{B_1}_{MF}$ ($\mathit{receivedTID}$) and the $PIN^{B_1}_{MF}$ ($receivedPIN$) of $B_1$ by checking whether these data exist in the $claimantAttributeList$. This list holds all the valid ownership claiming data of all the sold products whose ownership are yet to be claimed. If the $receivedPIN$ and $receivedTID$ are found in $claimantAttributeList$, the method returns an object, holding the data associated with the provided $TID^{B_1}_{MF}$ and $PIN^{B_1}_{MF}$. If the $claimantAttribute$ is invalid or empty, the request gets rejected. Then, $productCode$ is extracted from $claimantAttribute$ which is used to check in $productList$ that holds all the valid product data. If the associated $productData$ is not found, the request gets rejected, else the $generateVC$ function is invoked that generates $VC^{B_1}_{MF}$. Finally, the $VC^{B_1}_{MF}$ is returned.

The $generateVC$ function retrieves the credential definition ID ($credDefID$) used by $MF$ from their SSI wallet storage. Using the $credDefID$ and the product data($productData$), the $VC^{B_1}_{MF}$ is generated and returned. In future, if $B_1$ (Here, $B_1=S_1$; while selling the product to $B_2$) wants to transfer the ownership, $B_1$ (Here $B_1 = S_1$) first  exchange $TID^{B_2}_{S_1}$, and $PIN^{MF}_{B_2}$ with the expected buyer ($B_2$) and then sends a $ownershipTransferReq$ to $MF$. After that, $MF$ handles this request using the given $handleOwnershipTransferRequest$ function. First, the function checks whether the $productCode$ already exists in $claimantAttributeList$. If $isDuplicate$ is $TRUE$, this indicates the product is already being transferred to someone at that moment and therefore, the request gets rejected. Otherwise, the ownership claiming data such as $productCode$, $TID^{B_2}_{S_1}$, and $\{PIN^{MF}_{B_2}\}_{K_{{B_2}MF}}$ are stored in $claimantAttributeList$. Finally, the $ownershipProofReq$ is returned to $B_1$.

In Algorithm \ref{algo:handle-used-product-ownership-claim}, 
the $handleUsedProductOwnershipClaim$ function handles the $ownershipClaimReq$ sent by the buyer ($B_2$) to $MF$ to claim the ownership of a used product that $B_2$ has purchased from $S_1$ (here $S_1 = B_1$). $B_2$ exchanges $TID^{B_2}_{S_1}$, and $\{PIN^{MF}_{B_2}\}{K_{{B_2}{MF}}}$ with $S_1$ and $S_1$ forwards these data to $MF$. This function is invoked when $B_2$ sends the $ownershipClaimReq$ to $MF$. Now, within this function, the $TID^{B_2}_{S_1}$, and the symmetric key $K_{{B_2}{MF}}$ are extracted from the $ownershipClaimReq$. Next, the necessary $claimantAttribute$ data are filtered out by matching $TID^{B_2}_{S_1}$ and if the $claimantAttribute$ is not found, the $ownershipClaimReq$ gets rejected. Otherwise, a 3-4 digit random positive integer number denoted as $challengeBy$ and an arithmetic operation such as addition (+), subtraction (-), division (/), or multiplication (*) denoted as $challengeType$ gets generated. Then, the $challengeBy$, $challengeType$ and $key$ properties of the $claimantAttribute$ object get updated with the respective values. Next, the $pinChallengeReq$ object is generated and returned to $B_2$.

\begin{algorithm}
\SetAlgoLined
\caption{Algorithm Snippet for \texttt{handleUsedProductOwnershipClaim}, \texttt{handlePinChallengeRequest} \& \texttt{handlePinChallengeResponse} functions}
\label{algo:handle-used-product-ownership-claim}
\SetKwBlock{Begin}{}{}
\Begin(\textbf{Start})
{
  % ...
  \tcc{ The function below is handled by $MF$ when $B_2$ requests to claim ownership.}
    \SetKwFunction{handleUsedProductOwnership}{\textbf{handleUsedProductOwnershipClaim}}
      \SetKwProg{Fn}{function}{}{}
      \Fn{\handleUsedProductOwnership{$\mathit{ownershipClaimReq}$}}{        
        Extracts $TID^{B_2}_{S_1}$, and $K_{{B_2}{MF}}$ from the $ownershipClaimReq$ \\ 
        $\mathit{claimantAttribute}$ := $claimantAttributeList.matchAndReturn( TID^{B_2}_{S_1})$\;
            \lIf{ claimantAttribute is not found }{
                \KwRet reject $ownershipClaimReq$ 
            }
            \Else{
                Generate $challengeBy$, $challengeType$ data for challenging user PIN; \\
                Update $claimantAttribute$ with  $challengeBy$, $challengeType$ and key := $K_{{B_2}{MF}}$ in name-value pair \\ 
                Generate $pinChallengeReq$ object from $TID_{S_11}^{B_2}$ , $challengeBy$, $challengeType$ \\
                \KwRet $pinChallengeReq$;
            }
      }

\tcc{ The function below is handled by $B_2$ when $MF$ sends a  $pinChallengeReq$ to $B_2$. }
      \SetKwFunction{handlePinChallengeReq}{\textbf{handlePinChallengeRequest}}
      \SetKwProg{Fn}{function}{}{}
      \Fn{\handlePinChallengeReq{$\mathit{TID^{B_2}_{{S_1}}, challengeBy, challengeType}$}}{
        $\mathit{ownershipClaimingData}$ := Retrieve all the data received and shared with different sellers\;
        $\mathit{matchedData}$ := $ownershipClaimingData.matchAndReturn(TID^{B_2}_{S_1})$\;
            \lIf{ matchedData not found }{
                \KwRet reject $pinChallengeReq$
            }
            \Else{
             $expression$ := $generateExpression(matchedData.PIN^{MF}_{B_2}, challengeBy, challengeType )$ \\
             $challengeResult$ := $evaluate(expression)$ \\
              Generate $pinChallengeResp$ object from $TID_{S_11}^{B_2}$ and $challengeResult$ \\
            \KwRet $pinChallengeResp$
            }
      }

\tcc{ The function below is handled by $MF$ to respond with the $pinChallengeResp$ to $B_2$.}
      \SetKwFunction{handlePinChallengeResp}{\textbf{handlePinChallengeResponse}}
      \SetKwProg{Fn}{function}{}{}
      \Fn{\handlePinChallengeResp{$\mathit{ TID^{B_2}_{S_1}, challengeResult }$}}{
        $\mathit{claimantAttribute}$ := $claimantAttributeList$.$matchAndReturn( TID^{B_2}_{S_1} )$\;
            \lIf{ claimantAttribute not found }{
                \KwRet reject $pinChallengeResp$
            }
            \Else{
             $PIN^{MF}_{B_2}$ := $decryptPin(claimantAttribute.pin, claimantAttribute.key )$ \\
             $challengeValue$ := $claimantAttribute.challengeBy$ \\
             $operationType$ := $claimantAttribute.challengeType$ \\
             $expression$ := $generateExpression( PIN^{MF}_{B_2}, challengeValue, operationType )$ \\
             $mfResult$ := $evaluate(expression)$ \\

            \lIf{ $mfResult$ != $challengeResult$ }{
                \KwRet reject $pinChallengeResp$
            }
            \Else{            
                $newAttribute$ := $(connectionID, status, previouslySoldCount, lastPurchaseDate, email)$ \\
                $productData$ := $productList.matchAndReturn(claimantAttribute.productCode)$ \\
                Revoke the $VC^{B_1}_{MF}$ associated with $productData$ \\ 
                Update the $newAttribute$ of $productData$ and store in $productDataList$ \\ 
                \KwRet $revokeVC$ notification \\
            } 
            }
      }
  % ...
}
\end{algorithm}

After sending $pinChallengeReq$ to $B_2$, its different values are extracted and the $handlePinChallengeRequest$ function is invoked at the $B_2$'s end. Within the function, all data related to the transfer of ownership of different sellers are referred to as $ownershipClaimingData$. Next, the $matchAndReturn$ method checks if a match is found using $TID^{B_2}_{S_1}$, provided by $S_1$ earlier when $S_1$ sold the product to $B_2$. Upon a successful verification, the associated product data is returned. If there is no match, the request gets rejected. Otherwise, an arithmetic $expression$ is generated and  the $evaluate$ function computes the $challengeResult$ from the $expression$. Next, a $pinChallengeResp$ object is generated using $TID_{S_1}^{B_2}$ and the $challengeResult$ property. Finally, $pinChallengeResp$ is returned to $MF$. 

Once $MF$ receives $handlePinChallengeResponse$ from $B_2$, it extracts $TID^{B_2}_{S_1}$ and $challengeResult$ and invokes the $handlePinChallengeResponse$ function. First, it is checked if $TID^{B_2}_{S_1}$ is authentic by matching with $claimantAttributeList$. If not matched, the response gets rejected. Otherwise, the associated $PIN^{MF}_{B_2}$ of $claimantAttribute$ is decrypted using the respective key. Next, the values of $challengeBy$ and $challengeType$ are extracted from $claimantAttribute$ and an $expression$ is generated. This is exactly the same expression that $B_2$ used in $handlePinChallengeRequest$ function. Therefore, the $evaluate$ function evaluates the $expression$ and returns it to $mfResult$ variable. Now, if the $mfResult$ is not matched with $challengeResult$ received from $B_2$, the $pinChallengeResp$ gets rejected. Otherwise, $MF$ creates $newAttribute$ for the new owner and the current $productData$ is extracted using the $productCode$. The $newAttribute$ object holds some of the updated attribute of $productAttr$ for $B_2$ such as $ConnID$, $status$, $previouslySoldCount$, $lastPurchaseData$ and $email$. Next, the $VC^{B_1}_{MF}$ associated with $productData$ is revoked and the $productData$ is updated with $newAttribute$ and stored in $productDataList$. Finally, the $revokeVC$ notification is sent to $B_1$ to notify that $B_1$ is no longer the owner of the product and the VC is revoked.

\subsection{Use-case \& Protocol flow}
\label{subsection:usecase-and-protocol-flow}
In this section, we explore the use-cases of buying and claiming the ownership of a new product from a manufacturer and transferring that ownership as a seller to a new buyer. There are four scenarios through which the complete process is carried out. The first scenario is for buying the brand new product and the later three scenarios for transferring that ownership to a new buyer. Within these use-cases, the following activities take place: 

\begin{itemize}
    \item \textbf{Establishing the connection:} A secure P2P asynchronous SSI-based communication needs to be established between the system entities following the steps discussed in Section \ref{subsection:ssi}.
    
    \item \textbf{Providing VC to owners:} The Manufacturers should issue a VC to the owners that proves the ownership. This process is analogous to the traditional systems where either a paper-based document or credentials are provided as proof of ownership. 
    \item \textbf{Exchange ownership claiming attributes (PIN and Tracking ID):} While transferring or claiming ownership, buyers, sellers, and manufacturers exchange ownership claiming attributes such as PIN, TID and cryptographic key between themselves in a secure manner. These claiming attributes are required to claim ownership of the respective IoT device. This process is analogous to communication between a buyer and a seller before buying or selling a device in real life. However, here, the manufacturers are also involved to instill trust and authenticity to the whole mechanism. 
    \item \textbf{VC and ownership attribute storage:} The received VC and the ownership claiming attributes must be stored securely in the user's wallet. Additionally, manufacturers store these ownership attributes to maintain the ownership-claiming process.
    \item \textbf{Requesting proof:} The manufacturer requests the owner to provide proof of the issued VC using the same connection. This process is analogous to the authentication process in traditional web applications.
    \item \textbf{Revoke VC of owner:} While transferring the ownership of a device, the manufacturer first revokes the previously issued VC and issues a new VC to the new buyer.
\end{itemize}

While presenting these use-cases, we do not discuss the protocol to establish an SSI connection between two entities, as this is similar to any existing SSI use-cases, e.g. as presented in \cite{ferdous2022ssi4web}.

\subsubsection{Buying a brand new product:}
\label{subsubsection:buying-new-product}
In this use-case, we present the protocol flow for a buying a new IoT device by $B_1$ from a Distributor ($DS$, an authorised seller). The detailed protocol flow of this section is presented in Table \ref{table:protocol-buy-product}. Next, we present the protocol flow by denoting each of the steps as M[1.....n].  


\begin{table*}[h]
\caption{Buying Products from distributors}
\label{table:protocol-buy-product}
\centering
\begin{tabular}{lcrl}
\hline
$M1$ \quad $\mathit{DS} \rightarrow MF:$ ${[N_1, \mathit{productSellingReq}]}_{\mathit{HTTPS}}$ \\
$M2$ \quad $\mathit{MF} \rightarrow DS:$  ${[N_1, \mathit{productSellingResp}]}_{\mathit{HTTPS}}$ \\
$M3$  \quad $\mathit{MF} \rightarrow B_1:$  ${[N_1, \mathit{PIN_{MF}^{B_1}}]}_{\mathit{SMTP}}$ \\
\vspace{1.5mm} \noindent
$M4$ \quad $\mathit{DS} \rightarrow B_1:$  ${[N_1 \mathit{TID^{B_1}_{MF}}]}_{\mathit{SMTP}}$ \\ 
$M5$ \quad  ${ B_1 \mbox{ connects/re-connects with MF}}$ \\
$M6$ \quad $\mathit{B_1} \rightarrow MD:$ $[MF,\{N_2, \mathit{ownershipClaimReq(PIN^{B_1}_{MF}, TID^{B_1}_{MF}), \{ ownershipClaimReq \}_{K^{ -1 | MF}_{B_1}} }\}_{\mathit{K^{B_1}_{MF}}}]_{\mathit{K^{B_1}_{MD}}}$ \\
$M7$ \quad $MD \rightarrow MF:$ $[\{N_2, \mathit{ownershipClaimReq(PIN^{B_1}_{MF}, TID^{B_1}_{MF}), \{ ownershipClaimReq \}_{K^{ -1 | MF}_{B_1}} }\}_{\mathit{K^{B_1}_{MF}}}]_{\mathit{K^{MD}_{MF}}}$  \\
$M8$ \quad $MF \rightarrow MD:$ $[ B_1, \{N_2, \mathit{ownershipClaimResp, \{ ownershipClaimResp \}_{K^{-1 | B_1}_{MF}} }\}_{\mathit{K^{MF}_{B_1}}}]_{\mathit{K^{MF}_{MD}}}$  \\
$M9$ \quad $MD \rightarrow B_1:$ $[\{N_2, \mathit{ownershipClaimResp, \{ ownershipClaimResp \}_{K^{-1 | B_1}_{MF}} }\}_{\mathit{K^{MF}_{B_1}}}]_{\mathit{K^{MD}_{B_1}}}$  \\
$M10$ \quad $B_1 \rightarrow MD:$ $[MF, \{N_2, \mathit{ownershipClaimAck, \{ ownershipClaimAck \}_{K^{-1 | MF}_{B_1}} }\}_{\mathit{K^{B_1}_{MF}}}]_{\mathit{K^{B_1}_{MD}}}$  \\
$M11$ \quad $MD \rightarrow MF:$ $[\{N_2, \mathit{ownershipClaimAck, \{ ownershipClaimAck \}_{K^{-1 | MF}_{B_1}} }\}_{\mathit{K^{B_1}_{MF}}}]_{\mathit{K^{MD}_{MF}}}$  \\
\hline
\end{tabular}
\end{table*}

\begin{figure*} 
    \centering
  \subfloat[Claiming Brand New Product\label{subfig:sc_3_buy-new-claim}]{%
        \includegraphics[width=0.30\linewidth]{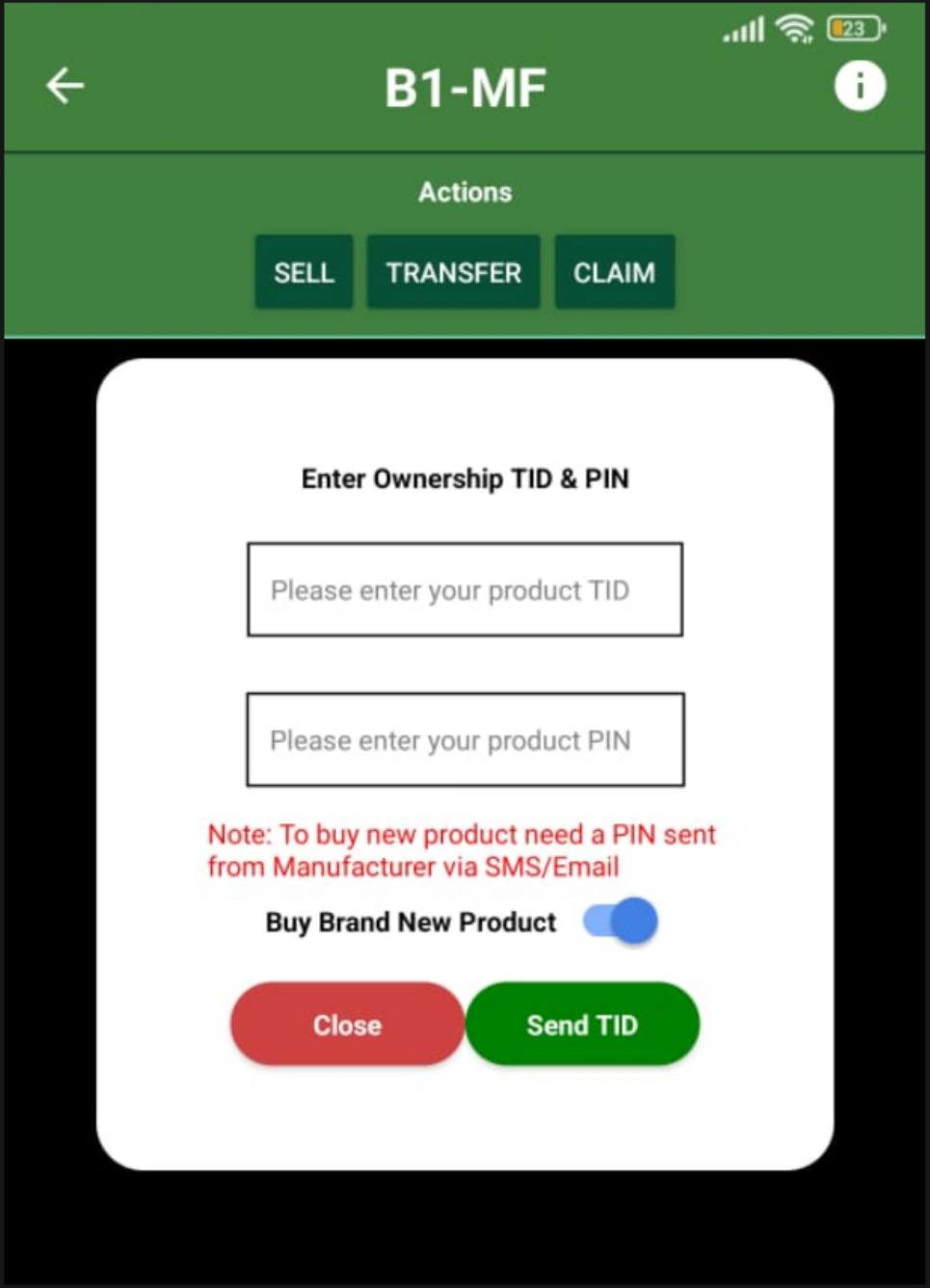}}
    \hfill
  \subfloat[$B_1$ receiving VC from $MF$ \label{subfig:sc_4_brand_new_vc}]{%
       \includegraphics[width=0.30\linewidth]{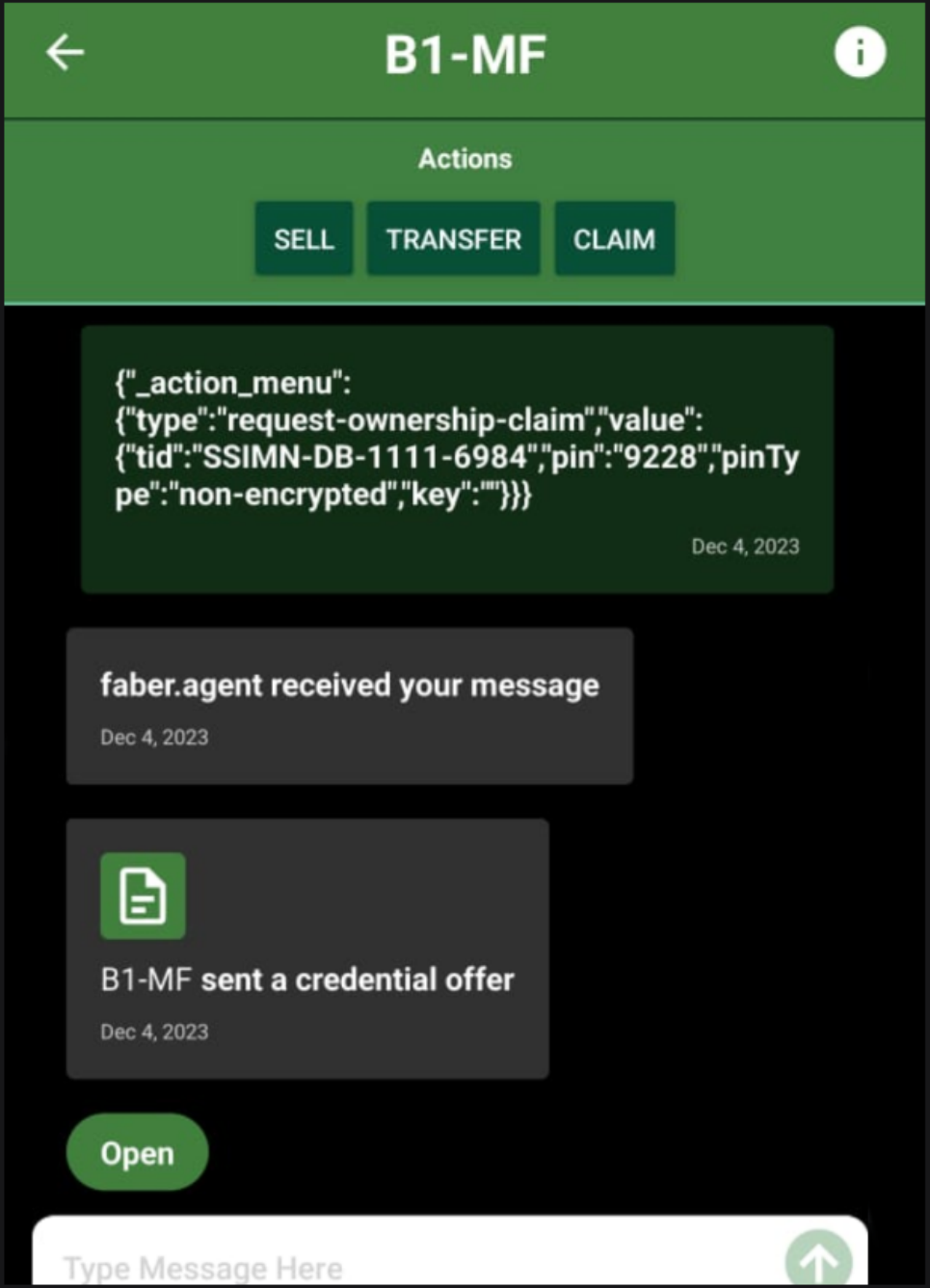}}
    \hfill
  \subfloat[$B_1$ reviewing VC attributes \label{subfig:sc_5_brand_new_vc_attr}]{%
        \includegraphics[width=0.30\linewidth]{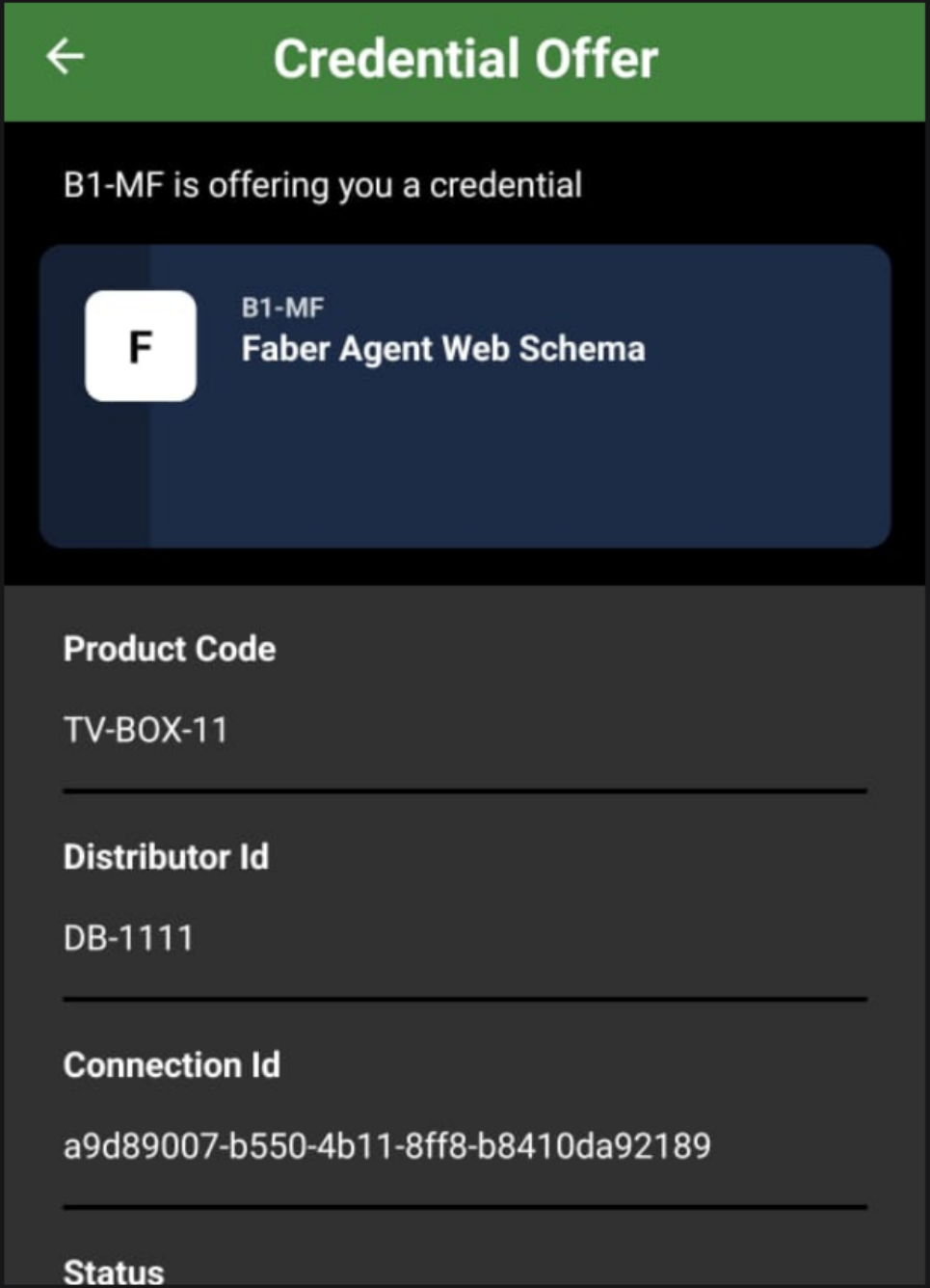}}
  \hfill
  \caption{Receiving, Reviewing and Accepting VC issued by $MF$ }
\end{figure*}

\begin{enumerate}[start=1,label={ \arabic* }]
    \item After selling a product to $B_1$ in person, a sales person from $DS$ enters the $productCode$ and $B_1$'s email on the $DS$'s web interface, facilitated by the manufacturer. The rest of the data for $productAttr$ are generated by the system. Next, these data are sent to the $MF$ (Manufacturer) by the $productSellingReq$ request (M1 in Table \ref{table:protocol-buy-product}). 
    \item $MF$ validates the information and returns a $productSellingResp$, consisting of $TID^{B_1}_{MF}$, to $DS$.
    \item $MF$ also sends a PIN ($PIN^{B_1}_{MF}$) to $B_1$ via email (M3 in Table \ref{table:protocol-buy-product}).
    \item When $DS$ receives $productSellingResp$, it extracts the $TID^{B_1}_{MF}$ from the response and sends it to $B_1$ via email (M4 in Table \ref{table:protocol-buy-product}).
    \item Now, $B_1$ uses the SSI wallet to scan a QR code to initiate an SSI connection with $MF$ ( M5 in Table \ref{table:protocol-buy-product}). For this, $B_1$ can use any computer if available or $DS$ can provide the QR code using a printed card or their computer. After getting connected with the $MF$, a messaging contacts option for $MF$ gets created in $B_1$'s mobile wallet application and to carry out rest of the steps with $MF$, $B_1$ utilises this contact option .
    \item Now, from the mobile wallet, $B_1$ selects the messaging option of the $MF$ from the contact list that bring a messaging box and from there, presses a "Claim" button that pops up a form, where $B_1$ actives the "Buy Brand New Product" toggle button and enters the $PIN^{B_1}_{MF}$ and $TID^{B_1}_{MF}$ received from $MF$ and $DS$ via email in steps M3 and M4 respectively (Figure \ref{subfig:sc_3_buy-new-claim}). After entering the prompted value, $B_1$ presses the "Send TID" button that sends the $ownershipClaimReq$, consisting of $PIN^{B_1}_{MF}$, $TID^{B_1}_{MF}$, and their associated signature ($\{ ownershipClaimReq\}_{K^{-1 | MF}_{B_1}}$), to $MD$ which is then forwarded the message to $MF$ (M6 and M7 in Table \ref{table:protocol-buy-product}).
    \item $MF$ extracts $PIN^{B_1}_{MF}$ and $TID^{B_1}_{MF}$ from $ownershipClaimReq$, verifies the signature using the $B_1$'s public key $K^{MF}_{B_1}$ and  validates the data of the associated product. If successful, $MF$ invokes the $\textit{handleNewProductOwnershipClaim}$ function of Algorithm \ref{algo:handle-new-product-ownership-claim} to send $VC^{B_1}_{MF}$ to $B_1$ via $MD$ within $ownershipClaimResp$ (M8 and M9 in Table \ref{table:protocol-buy-product}). 
    \item Once $B_1$ receives the $VC^{B_1}_{MF}$ from  in the message option (Figure \ref{subfig:sc_4_brand_new_vc}), $B_1$ opens it. The wallet verifies the signature of the VC using the public key of $MF$. If the verification is successful, $B_1$ can check the attributes of  $VC^{B_1}_{MF}$ (Figure \ref{subfig:sc_5_brand_new_vc_attr}). Upon clicking the "Accept" button, the $VC^{B_1}_{MF}$ is then stored in $B_1$'s wallet.
    \item After accepting the $VC^{B_1}_{MF}$, an acknowledgement response $ownershipClaimAck$ and its associated signature are sent to $MF$ via $MD$ that contains the acknowledgement of $B_1$'s action (M10 and M11 in Table \ref{table:protocol-buy-product}).  
\end{enumerate}
Now, $B_1$ has the $VC^{B_1}_{MF}$ as the proof of ownership of the IoT device.


\subsubsection{Exchanging a sensitive code between the current owner and a new buyer:}
\label{subsubsection:exchange-data-with-new-buyer}
Now, $B_1$ is the owner of the IoT device but wants to sell and transfer the ownership of the product to a new buyer $B_2$. This process requires three steps to complete. Since $B_1$ is currently selling the product, we imply $S_1 = B_1$. In this section, $S_1$ and $B_2$ exchange a secret between themselves that is necessary for the successful ownership transfer process. The protocol steps for this scenario are presented in Table \ref{table:protocol-ownership-transper-step-1} and discussed next.


\begin{table*}[h]
\caption{ Exchange TID and PIN between buyer($B_2$) and seller($S_1$); where $S_1=B_1$ }
\label{table:protocol-ownership-transper-step-1}
\centering
\begin{tabular}{lcrl}
\hline
$M1$ \quad  ${\mbox{An SSI Connection is established between \(S_1\) and \(B_2\) }}$ \\
$M2$ \quad $S_1 \rightarrow MD: [ B_2, \{N_1, PINReq, \{ PINReq\}_{K^{-1 | B_2}_{S_1}} \}_{\mathit{K^{S_1}_{B_2}}}]_{\mathit{K^{S_1}_{MD}}}$  \\
$M3$ \quad $MD \rightarrow B_2: [ \{N_1, PINReq, \{ PINRe\}_{K^{-1 | B_2}_{S_1}}  \}_{\mathit{K^{S_1}_{B_2}}}]_{\mathit{K^{MD}_{B_2}}}$  \\
$M4$ \quad $B_2 \rightarrow MD: [ S_1, \{ N_1, PINResp, \{ PINResp \}_{K^{-1 | S_1}_{B_2}}  \}_{\mathit{K^{B_2}_{S_1}}}]_{\mathit{K^{B_2}_{MD}}}$  \\
$M5$ \quad $MD \rightarrow S_1: [ \{N_1, PINResp,  \{ PINResp \}_{K^{-1 | S_1}_{B_2}}  \}_{\mathit{K^{B_2}_{S_1}}}]_{\mathit{K^{MD}_{S_1}}}$  \\
\hline
\end{tabular}
\end{table*}

\begin{figure} 
    \centering
  \subfloat[ Generating QR code to connect \label{subfig:sc_7_b1_qr_code}]{%
       \includegraphics[width=0.30\linewidth]{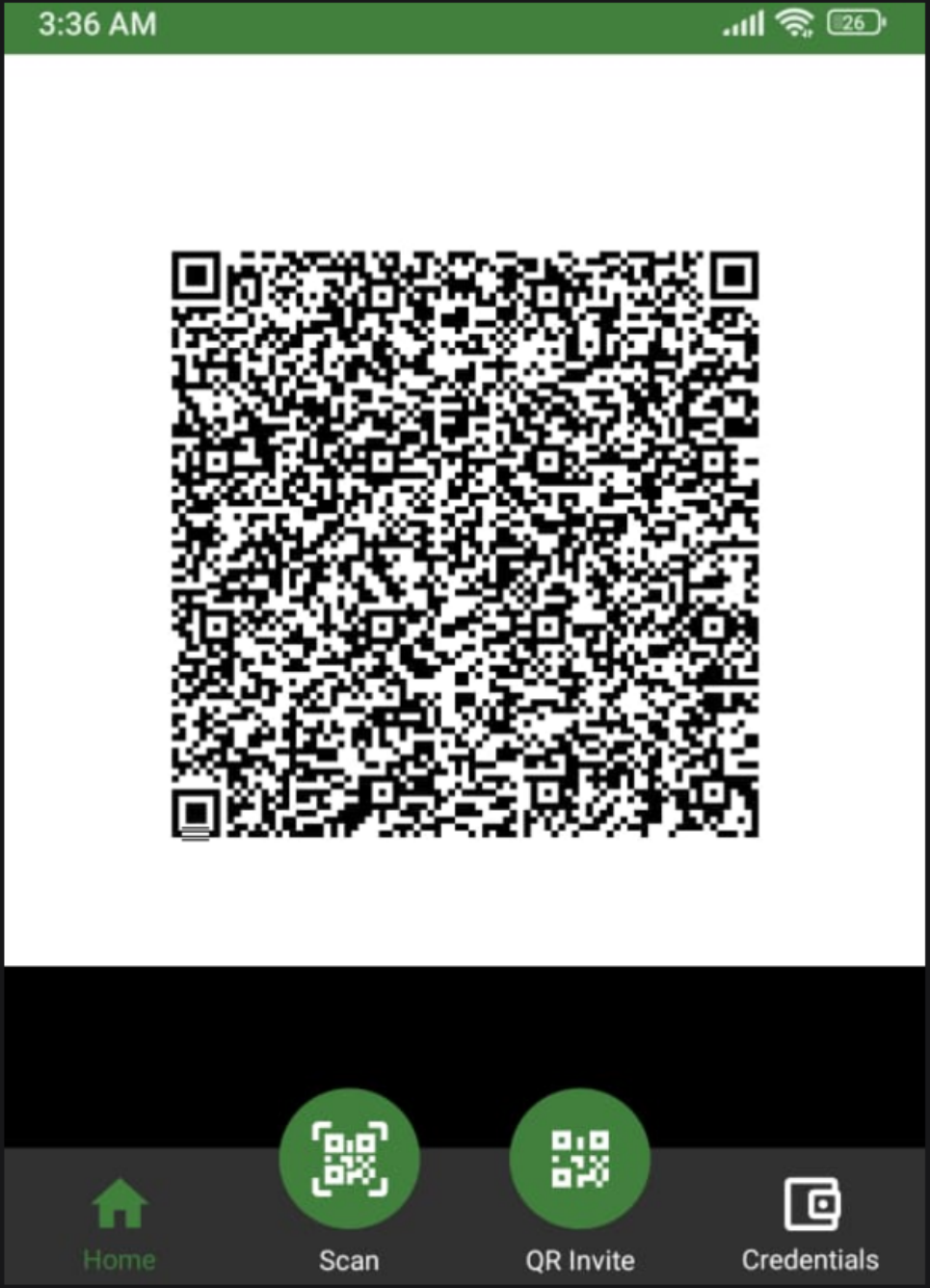}}
    \hfill
  \subfloat[$S_1$ sending pin request to $B_2$(My Wallet-4530)\label{subfig:sc_8_sell}]{%
        \includegraphics[width=0.30\linewidth]{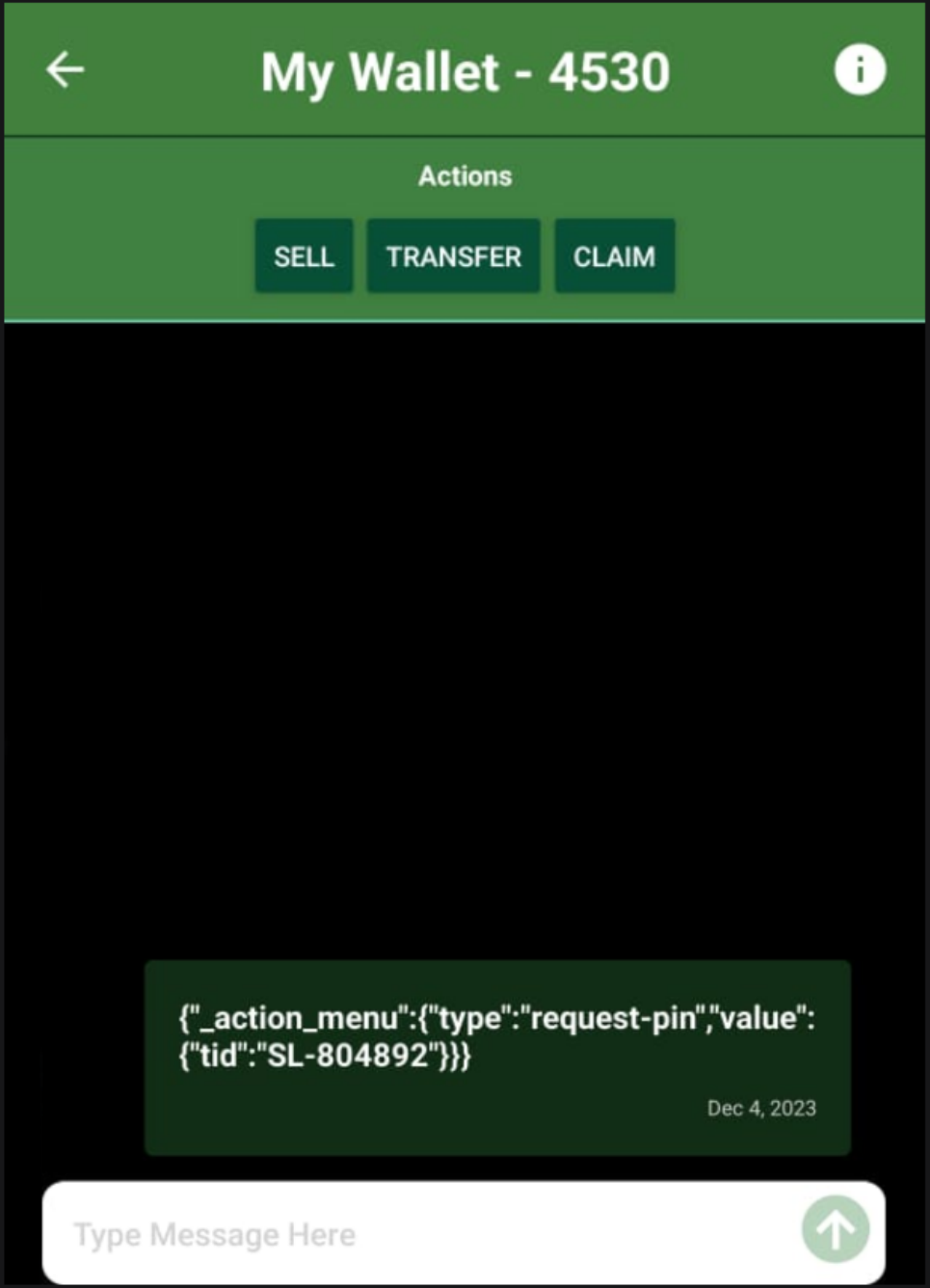}}
    \hfill
  \subfloat[$B_2$ sharing system generated pin with $S_1$(My Wallet - 864)\label{subfig:sc_9_b2_share_pin}]{%
        \includegraphics[width=0.30\linewidth]{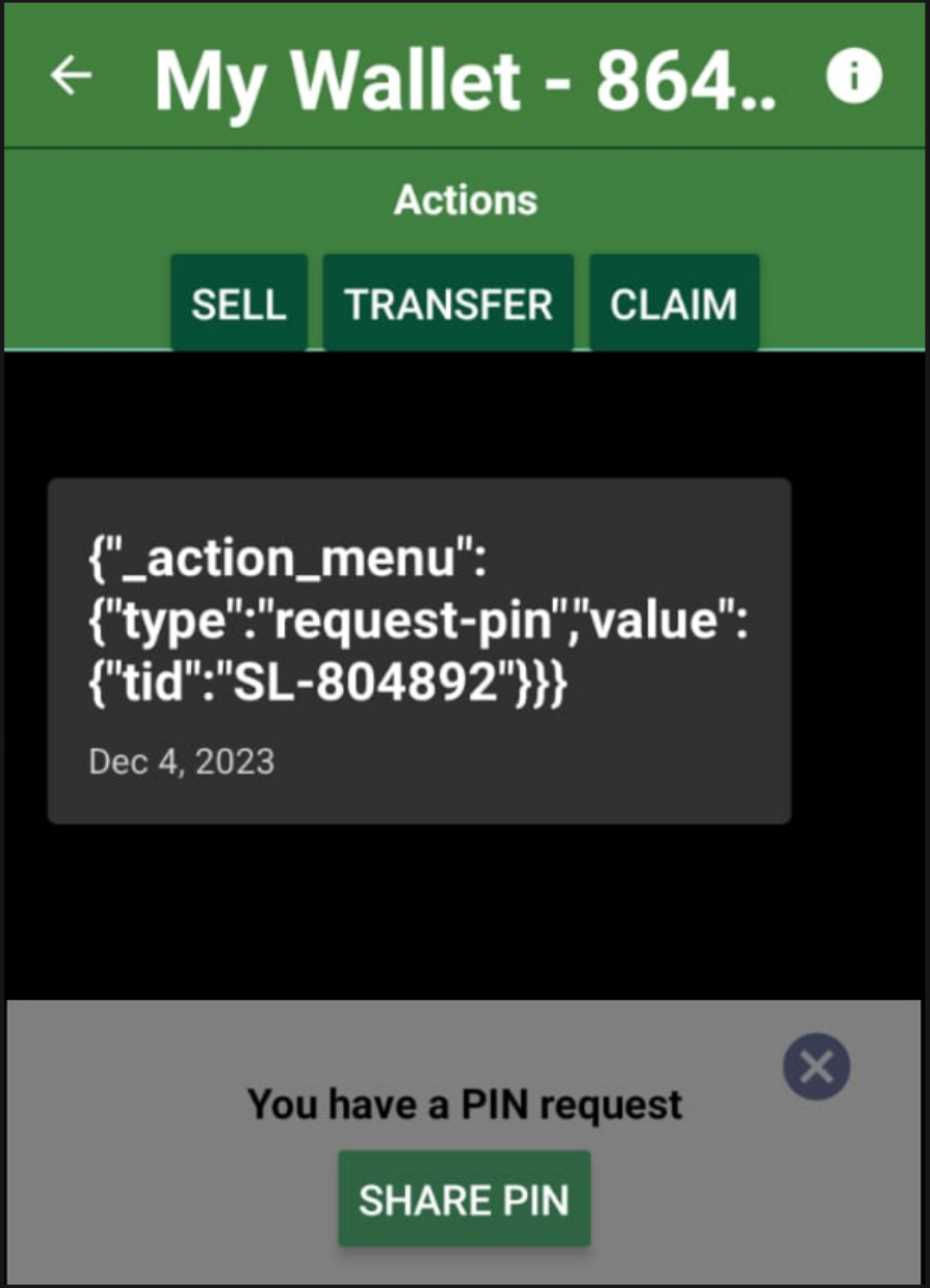}}
    \hfill
  \caption{ Connection establishment, TID and PIN exchanging between $S_1$ and $B_2$(where $S_1=B_1$).  }
  \vspace{-3mm}
\end{figure}

\begin{enumerate}[start=1,label={ \arabic* }]
    \item To exchange data, $S_1$ and $B_2$ must establish an SSI connection with each other. For this, either $S_1$ or $B_2$ generates a QR code from their mobile wallet by simply pressing the "QR Invite" button (Figure \ref{subfig:sc_7_b1_qr_code}), and the other scans it using the wallet to get connected (M1 in Table \ref{table:protocol-ownership-transper-step-1}). Once connected, a messaging contact option gets created in both of the user's mobile wallets. $S_1$ and $B_2$ utilise this contact option to communicate with each other securely. 
    \item Next, $S_1$ selects $B_2$ from the messaging contact option and presses the "Sell" button which generates $PINReq$, consisting of a freshly generated $TID^{B_2}_{S_1}$. The $PINReq$, its signature and a nonce ($N_1$) are encrypted using the public key of $B_2$ ($K^{S_1}_{B_2}$) and is sent to the $B_2$ via $MD$ (M2 and M3 in Table \ref{table:protocol-ownership-transper-step-1}).
    
    \item Once $B_2$ receives $PINReq$ and its signature, $B_2$ extracts the $TID^{B_2}_{S_1}$ and verifies its signature, and then stores them in the wallet, more specifically in $ownershipClaimingDataList$. In addition, a popup message with one button, "Share PIN", appears (Figure \ref{subfig:sc_9_b2_share_pin}). $B_2$ presses the "Share PIN" button which generates $PIN^{MF}_{B_2}$ and then encrypts it with a symmetric key ($K_{{B_2}{MF}}$). Next, the encrypted pin ($\{ PIN^{MF}_{B_2} \}_{K_{{B_2}{MF}}}$) and $TID^{B_2}_{S_1}$ are used to create $PINResp$ and $ \{ PINResp \}_{K^{-1 | S_1}_{B_2}}$, which are then sent to $S_1$ via $MD$ (M4 and M5 in Table \ref{table:protocol-ownership-transper-step-1}, Figure \ref{subfig:sc_10_b1_receive_enc_pin}). It is important to note that the encrypted pin ($\{ PIN^{MF}_{B_2} \}_{K_{{B_2}{MF}}}$) is generated by $B_2$ for $MF$ only. Therefore, the symmetric key to decrypt the PIN is not shared with $S_1$, rather $S_1$ just forwards this encrypted PIN to $MF$ when required (see below). Thus, only $B_2$ knows the actual PIN and the key.
\end{enumerate}

\subsubsection{Authorising the manufacturer to transfer the ownership:}
\label{subsubsection:authorization-to-transfer-ownership}
At this point, both $S_1$ and $B_2$ have $TID^{B_2}_{S_1}$ and the encrypted PIN ($\{ PIN^{MF}_{B_2} \}_{{B_2}{MF}}$). However, $S_1$ does not have the key to decrypt the PIN as it is only meant for $MF$. To authorise $MF$ to transfer the ownership, $S_1$ needs to prove the ownership and pass the encrypted pin and $TID^{B_2}_{S_1}$ to $MF$. The protocol flow of this process is presented in Table \ref{table:protocol-ownership-transper-step-2} and discussed next.

\begin{table}[h]
\caption{ $S_1$ authorises $MF$ to transfer the ownership of to $B_2$}
\label{table:protocol-ownership-transper-step-2}
\centering
\begin{tabular}{lcrl}
\hline
$M1$ \quad $S_1 \rightarrow MD: [ MF, \{ N_1, ownershipTransferReq, \{ ownershipTransferReq \}_{K^{-1 | MF}_{S_1}}  \}_{K^{S_1}_{MF}} ]_{K^{S_1}_{MD}}$  \\
$M2$ \quad $MD \rightarrow MF: [ \{ N_1, ownershipTransferReq, \{ ownershipTransferReq \}_{K^{-1 | MF}_{S_1}}  \}_{K^{S_1}_{MF}} ]_{K^{MD}_{MF}}$  \\
$M3$ \quad $MF \rightarrow MD: [ S_1, \{N_1, ownershipProofReq, \{ ownershipProofReq \}_{K^{-1 | S_1}_{MF}} \}_{ K^{MF}_{S_1}  }]_{ K^{MF}_{MD} }$  \\
$M4$ \quad $MD \rightarrow S_1: [\{N_1, ownershipProofReq, \{ ownershipProofReq \}_{K^{-1 | S_1}_{MF}}\}_{ K^{MF}_{S_1} }]_{ K^{MD}_{S_1} }$  \\
$M5$ \quad $S_1 \rightarrow MD: [MF, \{N_1, ownershipProofResp, \{ ownershipProofResp \}_{K^{-1 | MF}_{S_1}}\}_{ K^{S_1}_{MF} }]_{ K^{S_1}_{MD} }$  \\
$M6$ \quad $MD \rightarrow MF: [\{N_1, ownershipProofResp, \{ ownershipProofResp \}_{K^{-1 | MF}_{S_1}}\}_{ K^{S_1}_{MF}  }]_{ K^{MD}_{MF} }$  \\
$M7$ \quad $MF \rightarrow MD: [ S_1, \{ N_1, ownershipTransferResp, \{ ownershipTransferResp \}_{K^{-1 | S_1}_{MF}}  \}_{K^{MF}_{S_1}} ]_{K^{MF}_{MD}}$  \\
$M8$ \quad $MD \rightarrow S_1: [ \{ N_1, ownershipTransferResp, \{ ownershipTransferResp \}_{K^{-1 | S_1}_{MF}}  \}_{K^{MF}_{S_1}} ]_{K^{MD}_{S_1}}$  \\
\hline
\end{tabular}
\end{table}

\begin{figure*} 
    \centering
  \subfloat[$S_1$ received encrypted PIN from $B_2$'s wallet(My Wallet - 4530)\label{subfig:sc_10_b1_receive_enc_pin}]{%
       \includegraphics[width=0.30\linewidth]{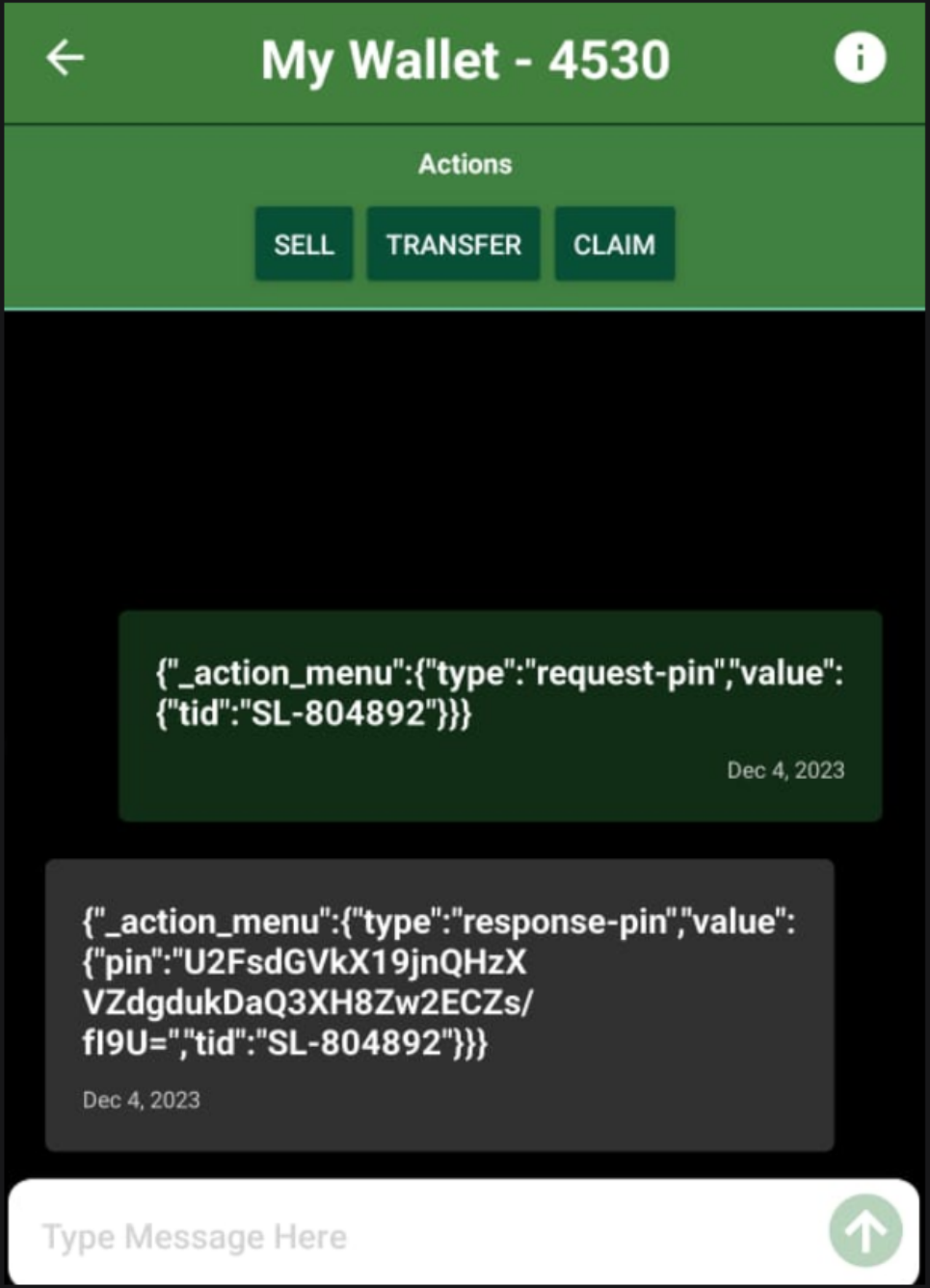}}
    \hfill
  \subfloat[$S_1$ Transferring TID, Product Code and encrypted PIN to $MF$\label{subfig:sc_11_transfer_req}]{%
        \includegraphics[width=0.30\linewidth]{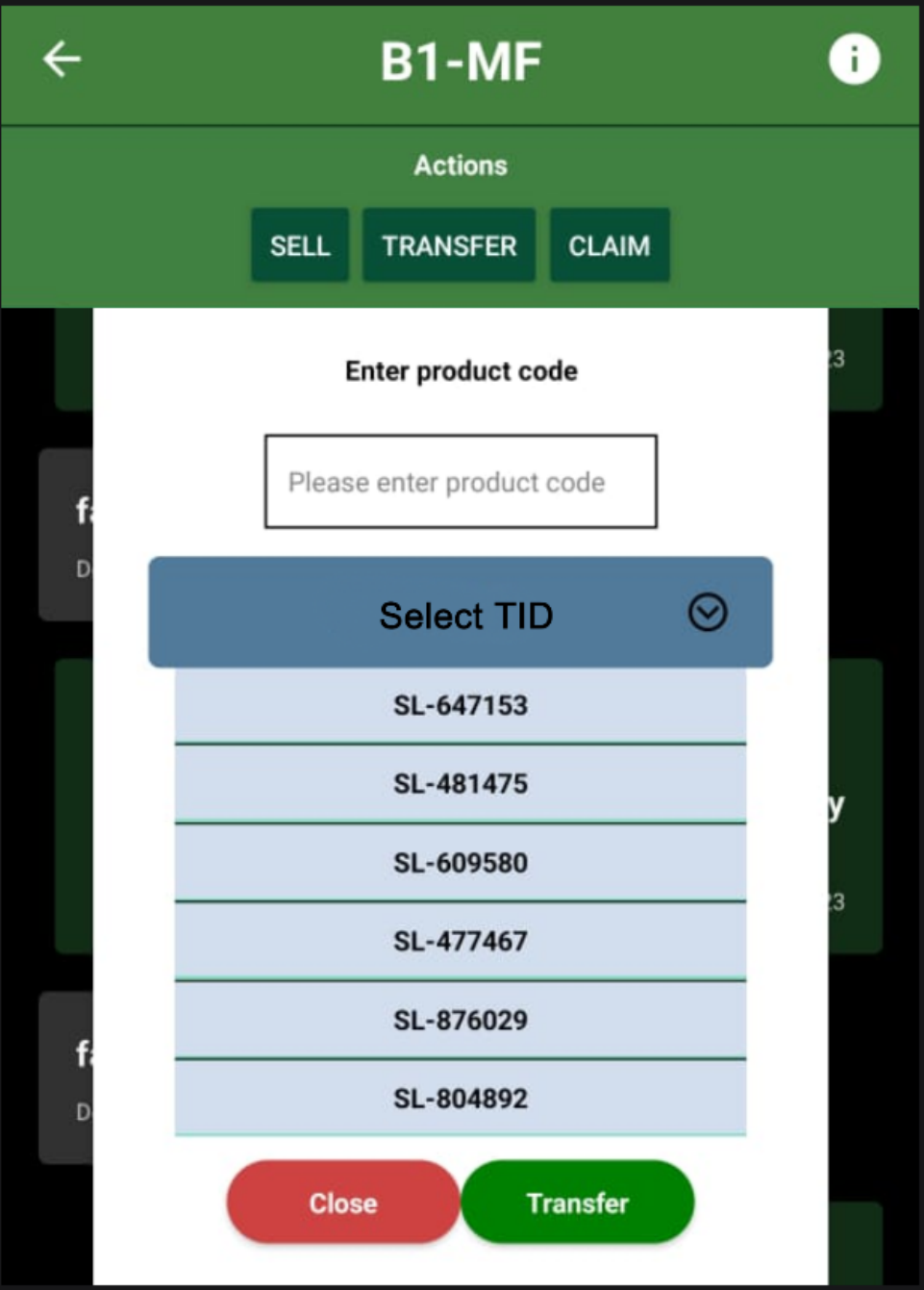}}
    \hfill
  \subfloat[ $B_2$ claiming ownership of $S_1$'s product from $MF$ \label{subfig:sc_12_b2_claim}]{%
        \includegraphics[width=0.30\linewidth]{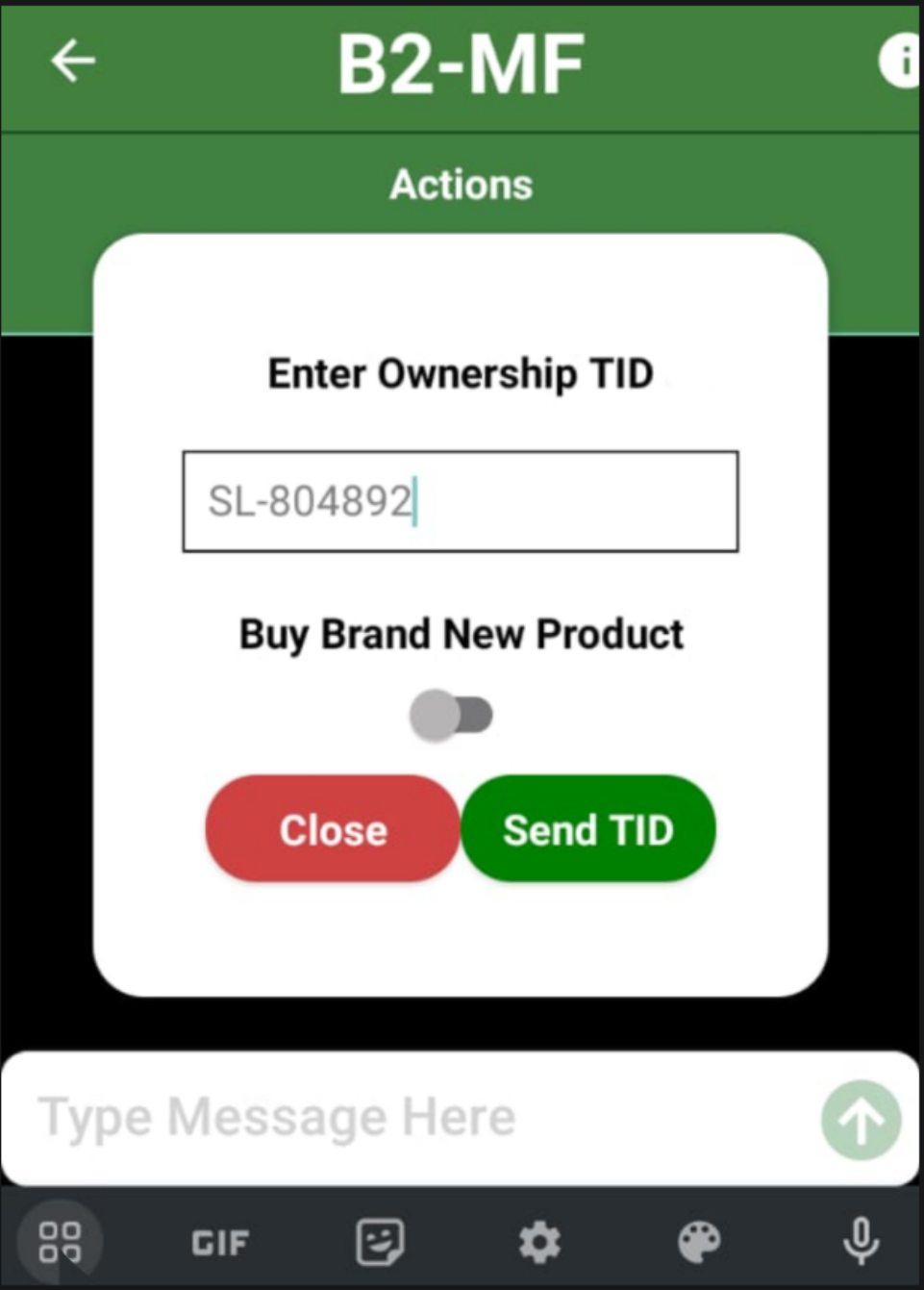}}
    \hfill
  \caption{ Exchanging product ownership claim data }
\end{figure*}

\begin{enumerate}[start=1,label={ \arabic* }]
     
    \item $S_1$ selects the $MF$ from the messaging contact list of the mobile wallet and press a button called "Transfer". This action shows a form where $S_1$ enters $productCode$ and $TID^{B_2}_{S_1}$ which were shared with $B_2$ (Figure \ref{subfig:sc_11_transfer_req}). The wallet fetches the associated encrypted pin $\{ {PIN^{MF}_{B_2}} \}_{K_{{B_2}MF}}$ from the $ownershipClaimingDataList$. Next, these three data are used to create $ownershipTransferReq$ and its associated signature, which are sent to $MF$ via $MD$ (M1 and M2 in Table \ref{table:protocol-ownership-transper-step-2}).    
    \item After receiving $ownershipTransferReq$, $MF$ checks the authenticity of the request by verifying the signature. Next, $MF$ wants to ensure that $S_1$ is the current valid owner and to do this, $MF$ sends $S_1$, via $MD$, a $ownershipProofReq$ (M3 and M4 in Table \ref{table:protocol-ownership-transper-step-2}).     
    \item When $S_1$ receives the $ownershipProofReq$ in the wallet, it selects the appropriate VC (${VC}^{B_1}_{MF}$) that satisfies the requested attributes encoded in $ownershipProofReq$. Next, $S_1$ presses the "Share" button to create an $ownershipProofResp$ with ${VC}^{B_1}_{MF}$ and some metadata and are sent to $MF$ including the associated signature via $MD$ (M5 and M6 in Table \ref{table:protocol-ownership-transper-step-2}).
    
    \item $MF$ extracts the VC from $ownershipProofResp$, verifies its signature to check the authenticity of the response and then validates the signature of the VC. Then attributes from the VC are extracted and used to verify whether the data are associated with a valid product. Afterwards $MF$ sends an $ownershipTrasferResp$ and the associated data signature back to $S_1$ via $MD$ (M7 and M8 in Table \ref{table:protocol-ownership-transper-step-2}).    
\end{enumerate}

At the end of this flow, $MF$ is now authorised to transfer the ownership of the associated product of $S_1$ to another probable owner $B_2$ only if $B_2$ can provide the necessary data to claim the ownership.


\subsubsection{Claiming the ownership of the used product:}
\label{subsubsection:claiming-ownership-of-used-product}
In this use-case, $B_2$ claims the ownership of the respective product from $MF$. In the previous step, $MF$ received the authorisation to transfer the ownership from $S_1$ to $B_2$. However, to claim this ownership, the prospective buyer $B_2$ needs to provide the $TID^{B_2}_{S_1}$ received from $S_1$ and the symmetric key $K_{{B_2}{MF}}$ to $MF$ so that $MF$ can decrypt the encrypted pin (${\{PIN^{MF}_{B_2}\} }_{K_{{B_2}MF}}$) and verify the claim. The protocol flows for this use-case are presented in Table \ref{table:claiming-new-ownership-of-used-product} and are discussed next.

\begin{table}[h]
\caption{Claiming the new ownership of used product from MF}
\label{table:claiming-new-ownership-of-used-product}
\centering
\begin{tabular}{lcrl}
\hline
$M1$ \quad $B_2 \rightarrow MD: [ MF, \{ N_1, ownershipClaimReq(TID^{B_2}_{S_1}, K_{{B_2}{MF}} ) , \{ ownershipClaimReq\})_{K^{-1 | MF}_{B_2}}\}^{B_2}_{MF}  ]_{K^{B_2}_{MD}}$  \\
$M2$ \quad $MD \rightarrow MF:$ $[ \{ N_1, ownershipClaimReq(TID^{B_2}_{S_1}, K_{{B_2}{MF}} ), \{ ownershipClaimReq \}_{K^{-1 | MF}_{B_2}}\}^{B_2}_{MF}  ]_{K^{MD}_{MF}}$  \\
$M3$ \quad $MF \rightarrow MD: [ B_2, \{ N_1, pinChallengeReq, \{ pinChallengeReq \}_{K^{-1 | B_2}_{MF}}\}^{MF}_{B_2}  ]_{K^{MF}_{MD}}$  \\
$M4$ \quad $MD \rightarrow B_2: [ \{ N_1, pinChallengeReq, \{ pinChallengeReq \}_{K^{-1 | B_2}_{MF}}\}^{MF}_{B_2}  ]_{K^{MD}_{B_2}}$  \\
$M5$ \quad $B_2 \rightarrow MD: [ MF, \{ N_1, pinChallengeResp, \{ pinChallengeResp \}_{K^{-1 | MF}_{B_2}}\}^{B_2}_{MF}  ]_{K^{B_2}_{MD}}$  \\
$M6$ \quad $MD \rightarrow MF: [ \{ N_1, pinChallengeResp, \{ pinChallengeResp \}_{K^{-1 | MF}_{B_2}}  \}^{B_2}_{MF}  ]_{K^{MD}_{MF}}$  \\
$M7$ \quad $MF \rightarrow MD: [S_1, \{N_1, revokeVC_{MF}^{B_1}, \{ revokeVC_{MF}^{B_1} \}_{K^{-1 | S_1}_{MF}}\}_{ K^{MF}_{S_1}   }]_{ K^{MF}_{MD}  }$  \\
$M8$ \quad $MD \rightarrow S_1: [\{N_1, revokeVC_{MF}^{B_1}, \{ revokeVC_{MF}^{B_1} \}_{K^{-1 | S_1}_{MF}}\}_{  K^{MF}_{S_1}  }]_{ K^{MD}_{S_1}  }$  \\
$M9$ \quad $S_1 \rightarrow MD: [MF, \{N_1, revokeVCResp, \{ revokeVCResp \}_{K^{-1 | MF}_{S_1}}\}_{  K^{S_1}_{MF} }]_{ K^{S_1}_{MD}} $  \\
$M10$ \quad $MD \rightarrow MF: [\{N_1, revokeVCResp, \{ revokeVCResp \}_{K^{-1 | MF}_{S_1}} \}_{  K^{S_1}_{MF} }]_{  K^{MD}_{MF}  }$  \\
$M11$ \quad $MF \rightarrow MD: [ B_2, \{N_2, ownershipClaimResp, \{ ownershipClaimResp \}_{K^{-1 | B_2}_{MF}}\}_{ K^{MF}_{B_2}  }]_{ K^{MF}_{MD}  }$  \\
$M12$ \quad $MD \rightarrow B_2: [\{N_2,  ownershipClaimResp, \{ ownershipClaimResp \}_{K^{-1 | B_2}_{MF}}\}_{ K^{MF}_{B_2}  }]_{ K^{MD}_{B_2}  }$  \\
$M13$ \quad $B_2 \rightarrow MD: [MF, \{N_2, ownershipClaimAck, \{ ownershipClaimAck \}_{K^{-1 | MF}_{B_2}}\}_{ K^{B_2}_{MF}  }]_{ K^{B_2}_{MD}  }$  \\
$M14$ \quad $MD \rightarrow MF:$ $[\{N_2, ownershipClaimAck, \{  ownershipClaimAck \}_{K^{-1 | MF}_{B_2}}\}_{ K^{B_2}_{MF}}  ]_{ K^{MD}_{MF}  }$  \\
\hline
\end{tabular}
\end{table}

\begin{figure*} 
    \centering
  \subfloat[$MF$ sends $pinChallengeReq$ to $B_2$\label{subfig:sc_14_pin_challenge_req}]{%
        \includegraphics[width=0.30\linewidth]{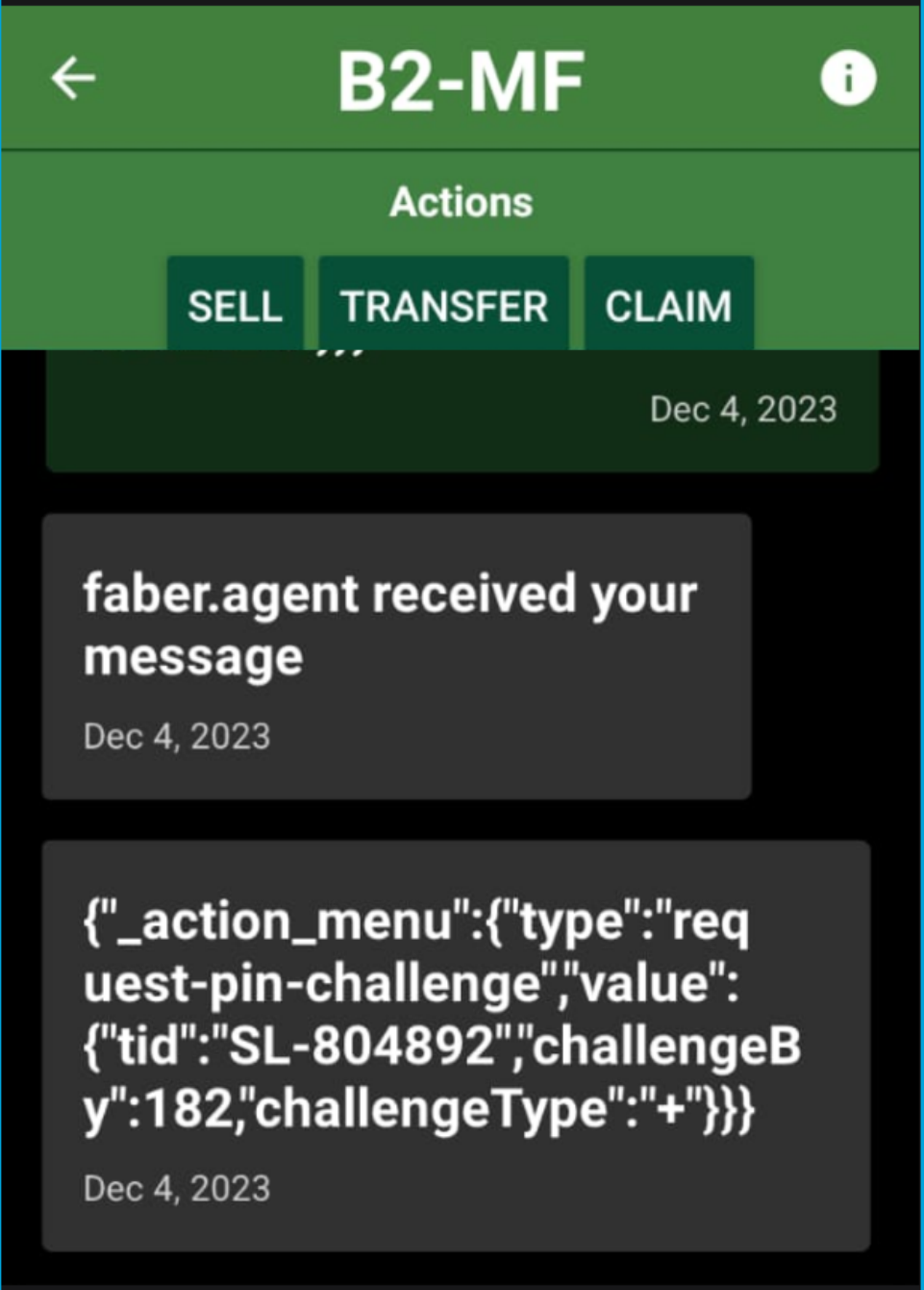}}
    \hfill
  \subfloat[$B_2$ replies with $pinChallengeResp$ and receives the VC of the product from $MF$. \label{subfig:sc_15_pin_challenge_resp}]{%
        \includegraphics[width=0.30\linewidth]{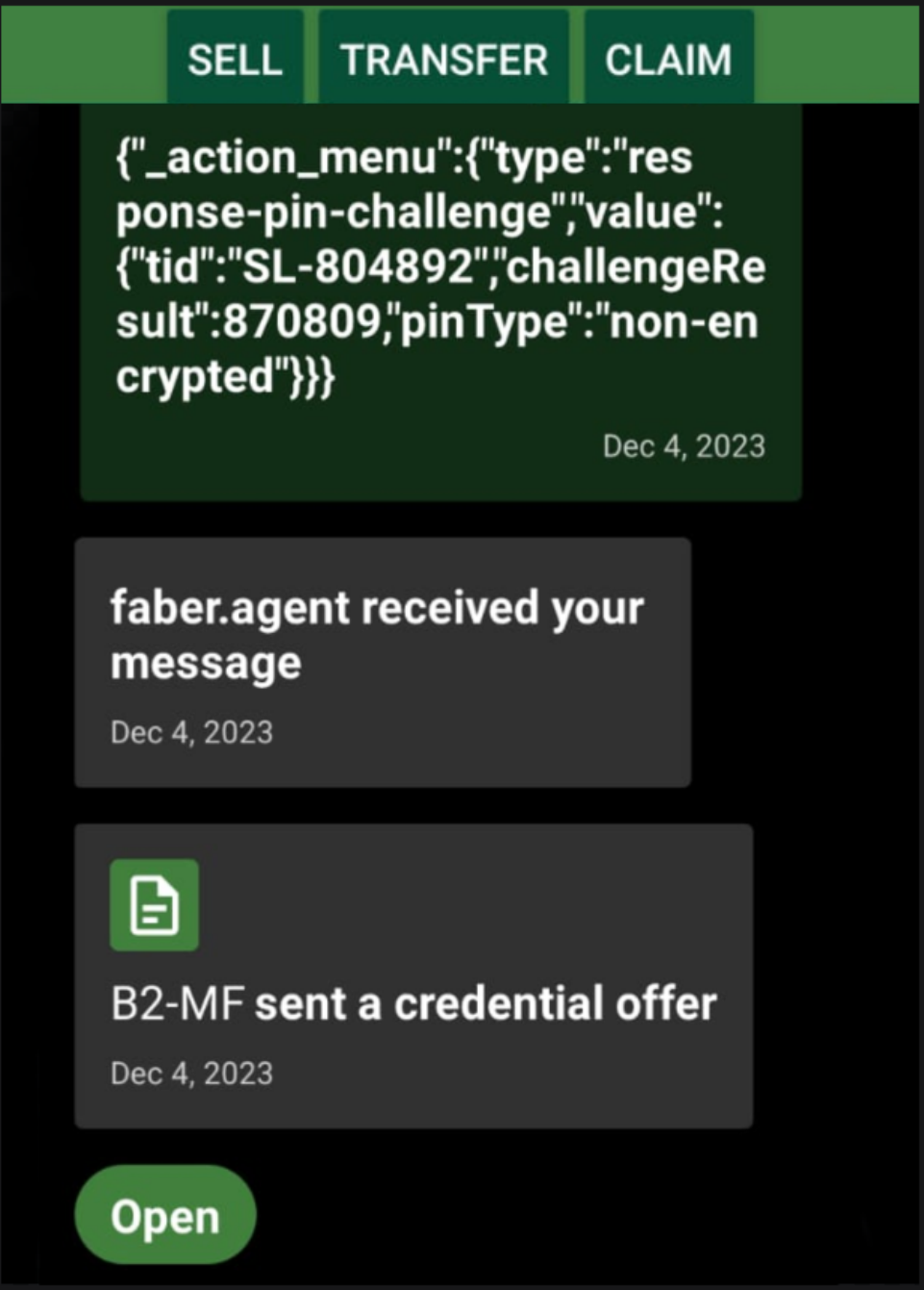}}
        \hfill
    \subfloat[$MF$ revoke VC of $B_1$ and send a notification \label{subfig:sc_15_revokevc}]{%
    \includegraphics[width=0.30\linewidth]{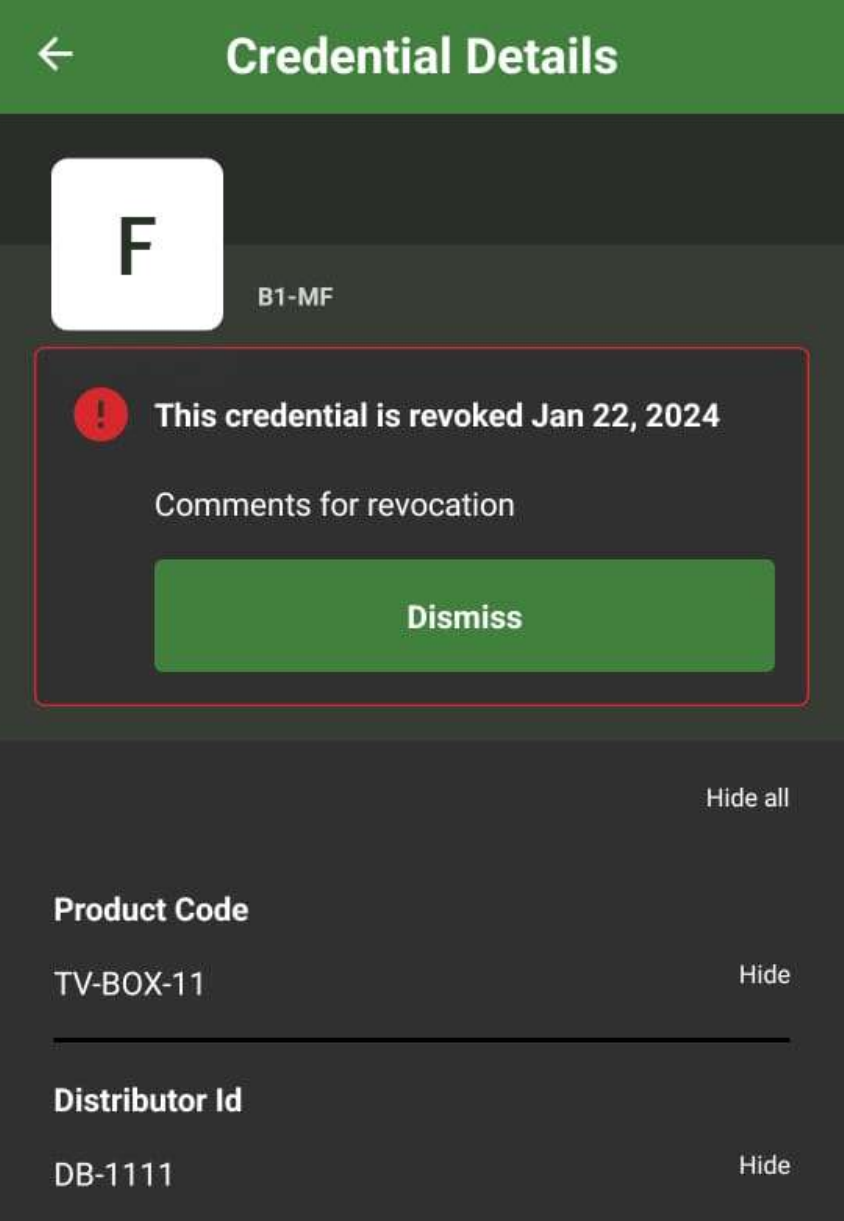}}
    \hfill 
  \caption{ $MF$ initiates pin challenge process and revoke VC }
\end{figure*}

\begin{enumerate}[start=1,label={ \arabic* }]    
    \item $B_2$ selects $MF$ from the contact list options and presses a "Claim" button which brings up an input field where $B_2$ enters the $TID^{B_2}_{S_1}$ received from $S_1$. After entering the value, $B_2$ presses the "Send TID" button which binds $TID^{B_2}_{S_1}$ and the associated symmetric key $K_{{B_2}{MF}}$ in $ownershipClaimReq$ along with the respective signature ($\{ownershipClaimReq\}_{K^{-1 | MF}_{B_2}}$), which are then sent to $MF$ through MD using the encrypted SSI connection (M1 and M2 in Table \ref{table:claiming-new-ownership-of-used-product}).   
    
    \item $MF$ invokes the $\mathit{handleUsedProductOwnershipClaim}$ of Algorithm \ref{algo:handle-used-product-ownership-claim} to create $pinChallengeReq$ along with a signature and is sent together to $B_1$ via $MD$ (M3 and M4 in Table \ref{table:claiming-new-ownership-of-used-product}). The $pinChallengeReq$ basically tells $B_2$ to modify the respective $PIN$ based on a system generated simple arithmetic operation. The expression of the arithmetic operation is generated from the values of "challengeBy" and "challengeType" attributes. Such attributes are shown in Figure \ref{subfig:sc_14_pin_challenge_req}. 
    
    \item The wallet of $B_2$ verifies signature $\{ pinChallengeReq \}_{K^{-1 | B_2}_{MF}}$ and invokes the $\mathit{handlePinChallengeRequest}$ function of Algorithm \ref{algo:handle-used-product-ownership-claim} to modify $PIN^{MF}_{B_2}$ as per the instructed operation in $pinChallengeReq$. After executing the operation, $B_2$ responds with $pinChallengeResp$ which holds the calculated result of modified PIN as the $challengeResult$ and its associated signature. This process is fully automated by the wallet without any human intervention from $B_2$. $pinChallengeResp$ is sent back to $MF$ via $MD$ (M5 and M6 in Table \ref{table:claiming-new-ownership-of-used-product}). 
    
    \item After receiving $pinChallengeResp$, $MF$ verifies the signature and invokes the $\mathit{handlePinChallengeResponse}$ function of Algorithm \ref{algo:handle-used-product-ownership-claim} to verify whether the requested operation on the PIN has been successfully executed. It is worth noting that $B_2$ has generated the encrypted PIN ($\{PIN_{B_2}^{MF}\}_{{B_2}{MF}}$ and shared this encrypted PIN with $MF$ via $S_1$ (while authorising $MF$ to transfer ownership, see in Section \ref{subsubsection:authorization-to-transfer-ownership}). Then, $B_2$ has shared the respective symmetric key via the encrypted SSI channel with $MF$ in step $M1$ of this protocol flow. This verification enables $MF$ to trust the claim of $B_2$ as the new owner and $S_1$ as the respective seller. After a successful verification, the previous ownership VC of $B_1$ (${VC^{B_1}_{MF}}$) is revoked and a notification is sent to $B_1$ via $MD$ (M7 and M8 in Table \ref{table:claiming-new-ownership-of-used-product}).     
    
    \item $S_1$ gets a notification in the wallet about the revocation of $VC^{B_1}_{MF}$ (Figure \ref{subfig:sc_15_revokevc}) and presses the "Ok" button to send a response ($revokeVCResp$) to $MF$ via $MD$ (M9 and M10 in Table \ref{table:claiming-new-ownership-of-used-product}). Consequently, the previous ownership VC of $S_1$ is revoked and $S_1$ cannot use it anymore.
    
    \item Now, $MF$ generates a new $VC^{B_2}_{MF}$ for $B_2$ and sends it to $B_1$ including the signature via $MD$ encoded in $ownershipClaimResp$ (M11 and M12 in Table \ref{table:claiming-new-ownership-of-used-product}).    
    
    \item After receiving the $ownershipClaimResp$ in the wallet, $B_2$ gets a notification of receiving a VC ($VC^{B_2}_{MF}$, Figure \ref{subfig:sc_15_pin_challenge_resp}). The wallet verifies the signature of the issued VC using the public key of $MF$ and stores it in the wallet. $B_2$ sends an acknowledgement message ($ownershipClaimAck$) to $MF$ via $MD$ (M13 and M14 in Table \ref{table:claiming-new-ownership-of-used-product}).    
\end{enumerate}

From now on, $B_2$ is the new valid owner of the product and can prove its ownership to anyone with $VC^{B_2}_{MF}$.


\section{\textbf{Performance Analysis}}
\label{section:performance-evaluation}
In our developed system, we have two key elements: an SSI agent-based 
manufacturer and the buyer or seller wallet, which is a native mobile application. Additionally, a simple distributor web application developed using Nodejs \cite{nodejs}. Since in the developed system users use their mobile wallet to manage and transfer their ownership through this mobile wallet, it is vital to analyse the performance of the wallet during different use-cases. Towards this aim, in this section, we discuss the configuration of our experimental setup for testing the wallet, followed by an analysis of its performance. The metrics that we have considered are CPU usage, memory consumption, crashes and energy consumption. 

\subsection{Experimental Setup and Performance Metrics}
\label{subsec:experimental-setup}
To conduct the experiment, we have used four distinct devices which are:
\begin{itemize}
  \item \textbf{Manufacturer and Distributor:} A laptop equipped with an Intel(R) Core(TM) i3-7100U CPU @ 2.40GHz, 12GB DDR4 RAM, 1TB HDD with 128GB SSD, and Ubuntu 22.0.04-64 OS. This device runs the SSI-based application of the Manufacturer and the web application of the Distributor.
  \item \textbf{User Wallets:} A Xiaomi Redmi Note 9s mobile device featuring an Octa-core 2.32GHz processor, Android 12, 6GB RAM, and 128 GB ROM. another mobile device, a Realme C11 with an Octa-core processor, Android 11, 4GB RAM, and 64GB ROM, has been used by the second user. However, we have evaluated the performance of the wallet running only the Xiaomi Redmi Note 9s.
  \item \textbf{Performance Observation:} To observe the performance, we have used Apptim \cite{apptim}, an open-source, user-friendly automated tool for analysing the performance of native mobile applications.
\end{itemize}

With this setup, we have simulated as if one user is managing the ownership of an IoT device and transferring the ownership to another user. While the user was using the Xiaomi Redmi Note 9s device, Apptim collected the performance metrics of the wallet and produced a comprehensive report of the test results which is discussed in the following section.

\subsection{Findings}
\label{subsection:findings}
In this section, we explore the findings of the experiment and discuss the observations we have from it.


The metrics collected by Apptim have three different ranges to define three different performance categories: \textit{Pass}, \textit{Moderate} and \textit{Warning}. The defined ranges and categories are presented in Table \ref{table:threshold}, which is based on Google's best practices and other market benchmarks \cite{appti-docs}.

Table \ref{table:threshold} also presents the result of the experiment for each of the metrics for the simulated use-case. We present a detailed discussion of the experimental result in the following. 

\begin{table}[H]
  \centering
    \begin{tabular}{|c|c|c|c|c|}
     
     \hline 
     Metric & Pass \textcolor{green}{\CIRCLE} & Moderate \textcolor{yellow}{\CIRCLE} & Warning \textcolor{red}{\CIRCLE} & Our Result  \\
    \hline 
     CPU Average (\%) & $<= 200$ & $> 200$ & $> 400$ & 39.3 (Pass \textcolor{green}{\CIRCLE} )  \\
     \hline
     CPU Max (\%) & $<= 200$ & $> 200$ & $> 400$ & 260.8 (Moderate \textcolor{yellow}{\CIRCLE} )  \\
     \hline
     Memory Average (MB) & $<= 256$ & $> 256$ & $> 400$ & 422.9 (Warning \textcolor{red}{\CIRCLE} )  \\
     \hline
     Memory Max (MB) & $<= 400$ & $> 400$ & $> 512$ & 463.1 (Moderate \textcolor{yellow}{\CIRCLE} )  \\
     \hline     
       Energy (pts) & $<= 250$ & $> 500$ & $> 1000$ & 131.3 (Pass \textcolor{green}{\CIRCLE} )  \\
     \hline
    \end{tabular}
  \caption{ Performance Summary }\label{table:threshold}
\end{table}

\begin{figure}[h]
\centering
\includegraphics[width=\textwidth]{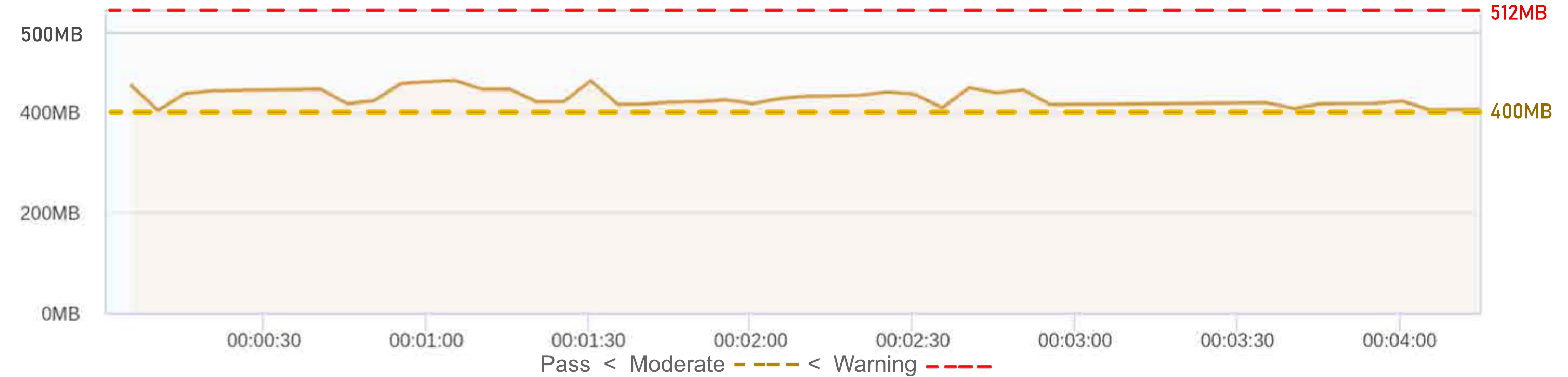}
\caption{App Memory Usage}
\label{fig:performance-app-memory}
\end{figure}

\begin{figure}[h]
\centering
\includegraphics[width=\textwidth]{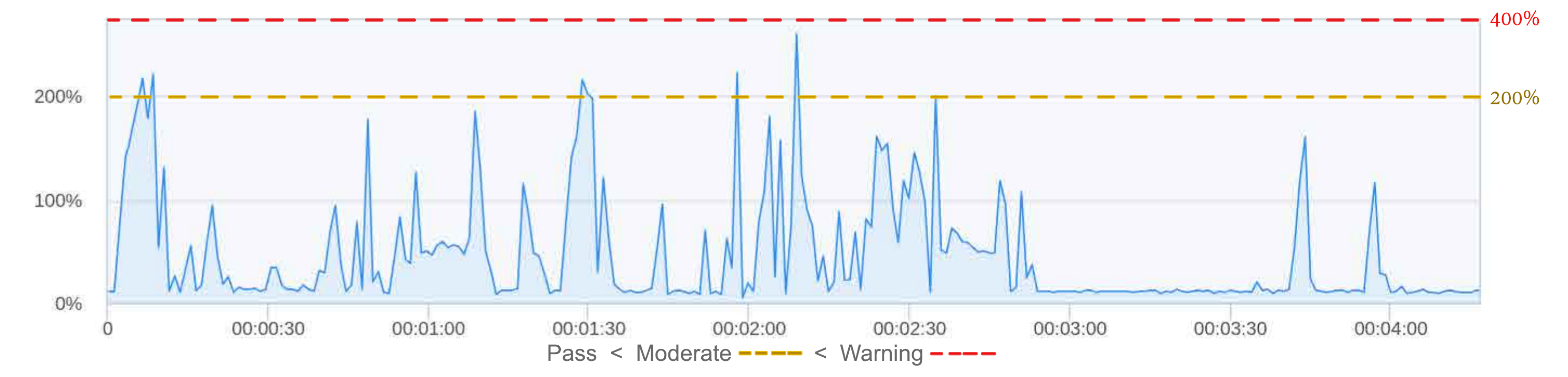}
\caption{App CPU Consumption}
\label{fig:performance-app-cpu}
\end{figure}

\noindent \textbf{Memory and CPU Usage:} Memory management is critical for any smartphone due to its memory-constrained nature that can impact the overall system performance. Figure \ref{fig:performance-app-memory} indicates that the memory usage by the SSI wallet mostly hovers around or above the 400MB threshold line, which is a moderate rate of \textit{maximum memory usage} range (Max usage > 512MB == warning). However, it can be seen from Table \ref{table:threshold} that the \textit{Average memory usage} surpasses the average warning value (Average usage > 400MB == warning). Additionally, efficient CPU utilisation is essential for any optimal mobile device performance. Our test result of CPU usage by the wallet (Figure \ref{fig:performance-app-cpu}) indicates a pass value of 39.3\%, significantly below the moderate threshold. However, there are several instances where the maximum app CPU usage surpasses the moderate threshold. One of the underlying reasons for this could be the execution of complex cryptographic operations that the wallet needs to carry out during different occasions. However, it is difficult to pinpoint the exact reason behind this behaviour.

\begin{figure}[h]
\centering
\includegraphics[width=\textwidth]{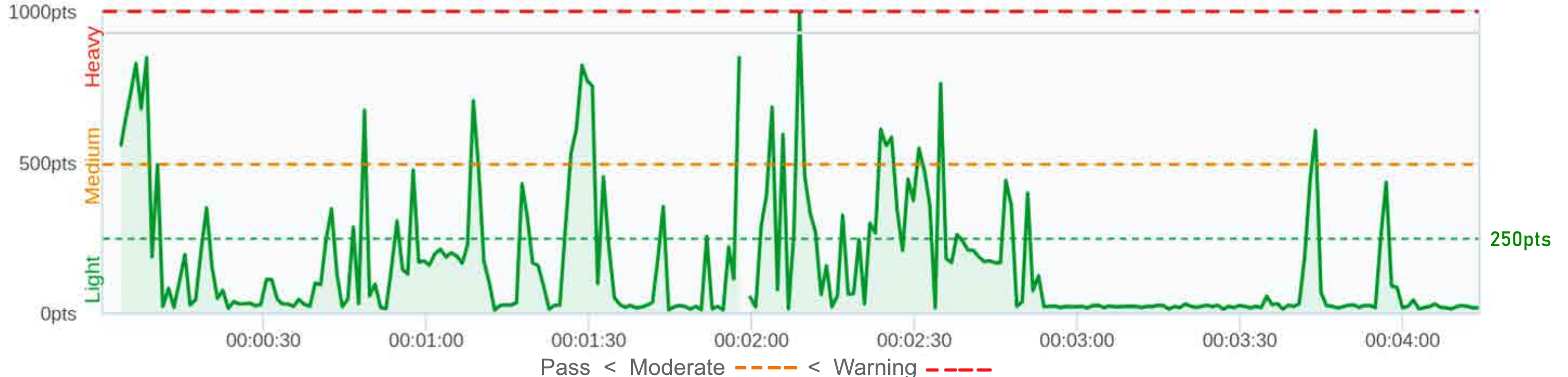}
\caption{Estimated Energy Consumption}
\label{fig:performance-energy}
\end{figure}

\noindent \textbf{Energy Consumption:} From Table \ref{table:threshold}, we see that the test shows a "Pass" status for the average energy consumption. However, there are several curves surpass the moderate line in Figure \ref{fig:performance-energy} and surprisingly the curve patterns and the timespan of these curves are similar to what we have seen for the CPU usage in Figure \ref{fig:performance-app-cpu}. We suspect these spikes are the result of higher CPU usage. 

\section{Discussion}
\label{section:discussion}
In this section, we analyse how the implemented PoC has satisfied different requirements (Section \ref{seb:reqAnalysis}), discuss its advantages and limitations (Section \ref{sec:ad}), present a comparative analysis between our proposed approach and existing research works (Section \ref{subsection:comparision-related-works}) and outline possible future works (Section \ref{subsec:fw}). 
\subsection{Analysing Requirements}
\label{seb:reqAnalysis}
\noindent \textbf{Functional Requirements:} The implementation of the proposed system satisfies all of the functional features that we discussed in section \ref{section:requirements-analysis}. From the protocol flow presented in Table \ref{table:protocol-buy-product}, it is evident that after emailing the PIN and TID by Manufacturer and Distributor respectively, all of the communications are done in a peer-to-peer manner where a mediator agent is used to transfer the data back-and-forth. This satisfies the requirement $FR1$. Also, from the protocol and the use-cases we see, owners use a wallet that holds their data in a form of VC. When manufacturers need to authenticate a user or users need to prove their ownership, the manufacturers verify the user’s identity information and ownership claim through this VC. In this process, other than the VC, the users do not require any other information such as password or passkeys. This satisfies the requirements $FR2$ and $FR3$. Moreover, when owners transfer their the ownership claim from the product to another user, the manufacturer first ensure that the current VC is revoked that makes the current owner unable to proof ownership of that particular product in future. Thus, the requirement $FR4$ is satisfied. 

\vspace{2mm}
\noindent \textbf{Security Requirements:} From the protocol flow(Table \ref{table:protocol-ownership-transper-step-2}), it is evident that when a user request for transferring their ownership, the manufacturer proves the user identity and ownership claim using their VC, that contains many data such as credential attributes, schema definition ID, credential definition ID, issuer's signature and public DID that helps the manufacturer to identify the issuing entity and the legality of the VC. Additionally, when a new buyer request to claim VC for a product he/she has bought from another user, the manufacturer sends a pin challenge and in return the owner responds with a symmetric key and result associated with the PIN. The Manufacturer validates the result and the symmetric key to ensure the prospective owner is the real one. In addition, the use of memorable name while creating connection (as proposed by Ferdous et. al. in \cite{ferdous2022ssi4web}) prevents that the malicious users cannot use others connection id. This satisfies the security requirement $SR1$. The VCs are signed by the manufacturers and therefore and hence, any tampering will be immediately identified and rejected. Additionally, the use of digital signature at different stages of the protocol, sharing of the encrypted PIN and the symmetric key ensure that both $SR2$ and $SR3$ are satisfied. It is evident from the protocol that all of the communications and data are encrypted and the system works in a request-response manner that convince the $SR4$. To mitigate $SR5$, multiple precautionary measures should be taken, since initiating a DoS attack does not require any vulnerabilities in the application or network. This attack can be target to any web services and therefore, as precautionary step to mitigate this threat, the application developers are required to adopt different mechanisms while implementing the , which is not in the scope of our research. For $SR6$, it is required to adopt and deploy an access control mechanism and this is more relevant when some services can be used by users with proper authorization. Since, we are not working to use the features of devices or the services, this threat is not taken into consideration in our work. To mitigate the $SR7$, nonces have been used at all steps in each use-case. Finally, to support our claim about the satisfaction of security requirements, we have validated the protocol using $ProVerif$.

\vspace{2mm}
\noindent \textbf{Privacy Requirements:} In the proposed protocol, the users first send a claim request to manufacturers before the manufacturers issue VC to any user. Also, the valid owners must request the manufacturers to transfer their ownership to a new user. This satisfies the $PR1$.

\subsection{Protocol Validation}
\label{sec:pv}

In order to assess the system's security, we conduct a formal verification of the protocol using an advanced protocol verification tool, ProVerif \cite{blanchet2018proverif}. Our primary focus lies on the secrecy and authentication goals of the protocol. As an initial step, we have formalized the protocol with ProVerif, so that it can analyse the protocol's security attributes and detect any possible vulnerabilities. Following that, we have executed a model-checking procedure that ascertains whether the system satisfies the defined security properties. This has allowed us to confirm if our protocol achieved the necessary secrecy and authentication goals, thereby confirming the system's overall security.

The ProVerif scripts for each protocol can be accessed from the link in the footnote \footnote{\url{https://drive.google.com/uc?export=download&id=1eqEpwItVj0Pvp4kVmGpWiZHc5dzSvQeT}}. In the following, we discuss the secrecy and authenticity goals of these ProVerif scripts.

\begin{figure*}[htbp]
\centering
\includegraphics[width=\textwidth]{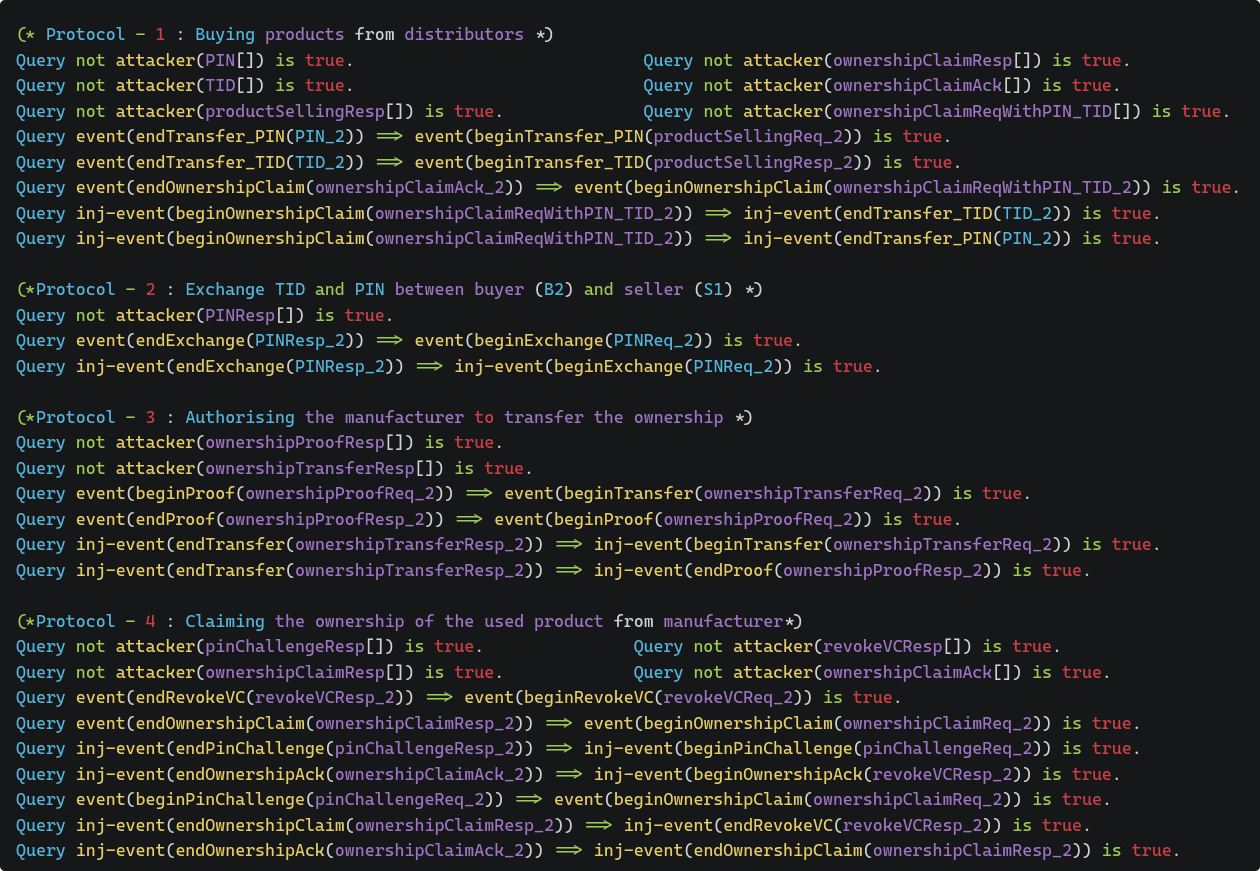}
\caption{ProVerif scripts results}
\label{fig:proverif_results}
\end{figure*}

The secrecy and authenticity goals outline the elements of confidential data that are vulnerable to unauthorised disclosure or modification by a malicious entity. Our ProVerif analysis of the protocol has demonstrated robust safeguarding of all specified secrecy properties against potential adversarial threats. Moreover, the protocol's design effectively prevents any entity from disrupting the critical event sequence, thereby ensuring the integrity and authenticity of the transactional flow. ProVerif can display three kinds of results which are RESULT [Query] is true, false, or cannot be proved. After executing the ProVerif scripts we have received true outputs for all queries. Figure \ref{fig:proverif_results} shows the results of the four ProVerif scripts corresponding to our proposed system protocols sequentially. This result provides a proof that both the secrecy goals and the authenticity goals are valid, thus satisfying the secrecy and authenticity goals. 

\subsection{Advantages and Limitations}
\label{sec:ad}
The advantages of SSI-based ownership management and transfer system are:
\begin{itemize}
\item The system is completely passwordless. Therefore, users do not need to use any type of password to manage the ownership of their smart appliances. 
\item Aside from email addresses, no personal information from users is retained in the manufacturer's database. This is useful for users. This is also beneficial for manufacturers as they do not have to deal with more and more sensitive user data. 
\item This approach also eases after-sales service for the first owner as well as for any subsequent owner.
\item  Control over IoT devices will be more secure and easier.
\end{itemize}

There is one particular challenge of the presented solution. Since it is holistic in nature and would require to make significant changes to the existing infrastructure. This will impact every single entity in the ecosystem. Therefore, its adoption might be challenging if all entities are not on-boarded seamlessly.

\subsection{Comparative analysis with existing related research works}
\label{subsection:comparision-related-works}

In this section, we present a comparative analysis between existing related research works and our approach. Table \ref{table:comparison-related-work} presents the result of the comparative analysis. We have employed various properties for the comparison, all of which are self-explanatory. Inside the table, the `\CIRCLE' symbol is used to signify that a certain property is satisfied by the respective research, the `\RIGHTcircle' signifies that a certain property is partially satisfied by the respective work and lastly, the `\Circle' symbol signifies that the research does not satisfy the property. Table \ref{table:comparison-related-work} clearly highlights the differences between our SSI based approach and other related works. It is evident that this work is the first to propose an ownership transfer mechanism using a passwordless mechanism with a detailed protocol. The protocol is validated to ensure that the security requirements are fulfilled. The use of SSI ensures that this approach is user-centric, with users having ultimate control over how the ownership of their smart appliances can be securely managed.   

\begin{table}[!h]
  \centering
    \begin{tabular}{ p{20mm}|p{12mm}|p{12mm}|p{10mm}|p{12mm}|p{13mm}|p{12mm}|p{13mm}|p{16mm}}
     \hline
     &  B.A. Barakati et al.  \cite{HC1} & M. Alblooshi et al. \cite{HC2} & V. Aski el al.\cite{sc1} & M. Gunnarsson el al. \cite{sc2} & M.S.N. Khan et al. \cite{sc4} & S.F. Aghili et al. \cite{sc5} & Kiran M. et. al.\cite{New-1}  & Our Approach \\ 
     \hline
     \hline 
      Threat Model                                           & \Circle  & \Circle  & Dolev-Yao (\RIGHTcircle) & Dolev-Yao (\CIRCLE) & \Circle & Dolev-Yao (\CIRCLE) & \Circle & STRIDE (\CIRCLE) \\
     \hline
      Security Requirements                                      & \Circle &  \Circle  &  \CIRCLE  &  \CIRCLE  &   \CIRCLE  &  \CIRCLE & \Circle  & \CIRCLE \\
        \hline
      Functional Requirement                                & \CIRCLE & \Circle &  \CIRCLE & \Circle & \CIRCLE &  \CIRCLE & \Circle & \CIRCLE \\
     \hline
      Security Analysis                                      & \CIRCLE &  \CIRCLE  &  \CIRCLE  &  \CIRCLE  &   \CIRCLE  &  \CIRCLE  & \CIRCLE & \CIRCLE \\
     \hline
     Protocol Design                                         & \Circle & \Circle & \Circle & \CIRCLE & \Circle & \CIRCLE & \CIRCLE & \CIRCLE \\
     \hline
     Protocol Validation                                    & \Circle & \Circle & \Circle & \CIRCLE & \Circle & \CIRCLE & \CIRCLE & \CIRCLE \\ 
     \hline
     Implementation                                   & \CIRCLE & \RIGHTcircle & \Circle & \CIRCLE & \CIRCLE & \Circle & \CIRCLE & \CIRCLE \\
     \hline
     Core technology utilised                              & Ethereum \cite{nodejs} & Ethereum & \Circle & Contiki-NG (OS) \cite{Contiki-NG-os} & Respburry-Pi \cite{raspberry-pi}, Android \cite{android} & \Circle & Ethereum & Hyperledger Indy, Hyperledger Aries, Aries Bifold Wallet \\
     \hline
     Performance Analysis                                  & \Circle & \Circle & \Circle & \CIRCLE & \CIRCLE & \Circle & \Circle & \CIRCLE \\
     \hline
     Dependency                                              & \Circle & \Circle & Smart card (\CIRCLE) & Trusted Third Party (\CIRCLE) & \Circle & Smart card (\CIRCLE) & \Circle & \Circle \\
     \hline
     User-centric                                            & \Circle & \Circle & \Circle & \Circle & \Circle & \Circle & \Circle & \CIRCLE \\
     \hline
     \hline
    \end{tabular}
  \caption{ Comparison with the related research on ownership transfer }\label{table:comparison-related-work}
\end{table}

\subsection{Future Work}
\label{subsec:fw}

\begin{itemize}

\item If the user wallet is lost, its stored VCs are lost too. Therefore, it is important to safely backup and restore any VC from the wallet so that even if the wallet gets lost, any VC can be recovered. Authors in \cite{farhad2024secure} presented a secure backup and restore mechanism for Aries based Wallet. In future, we will add this feature in our wallet.
\item The presented solution has been designed for smart-appliances. In future, we will explore how this can be integrated with other IoT devices as well.
\item The presented solution offers a new way for ownership management of consumer IoT devices. The mechanism is a stark opposite to the current approach. In addition, the concept of SSI is comparatively new. In order to ensure a wide-scale adoption of this approach, it is important to conduct a usability study of the proposed system. In future, we would like to conduct a thorough usability study.
\end{itemize}

\section{Conclusion}
\label{section:conclusion}
In this work, we have presented an SSI-based ownership management and transfer system for consumer IoT devices. SSI technology makes our system completely passwordless and self-sovereign, where the user is the center of the system. Without the permission of the user, no activities for ownership management and transfer are possible. The motivation is that users should have the ultimate control of the ownership data and only with their explicit involvement and consent, the ownership of a smart appliance be transferred to another user. We have designed the architecture of the system, based on a threat model and requirement analysis, discussed its implementation details and highlighted multiple use-cases to show its applicability. We have also analysed the performance of developed system and formally verified its security using ProVerif, a well-known protocol verification tool. The proposed system presents a holistic solution, which has the potential to open up new doors for research within different IoT application domains. 

\bibliographystyle{ACM-Reference-Format}
\bibliography{main}
\end{document}